\documentclass[12pt]{article}

\catcode`\@=11

\global\arraycolsep=2pt
\usepackage{amsbsy,amssymb,latexsym,amsfonts,amsmath}
\usepackage{graphicx,color}

\allowdisplaybreaks

\begin{document}

\begin{flushright}
RUP-24-2\\
YITP-24-11

\end{flushright}

\vspace*{0.7cm}

\begin{center}
{ \Large  Convexity restoration from hairy black hole in Einstein-Maxwell-charged scalar system in AdS}
\vspace*{1.5cm}\\
{Takaaki Ishii and Yu Nakayama}
\end{center}
\vspace*{1.0cm}
\begin{center}

Department of Physics, Rikkyo University, Nishi-Ikebukuro, Toshima-ku, Tokyo~171-8501, Japan\\
Yukawa Institute for Theoretical Physics,
Kyoto University, Kitashirakawa Oiwakecho, Sakyo-ku, Kyoto 606-8502, Japan

\vspace{1.5cm}
\end{center}

\begin{abstract}
In the Einstein-Maxwell-charged scalar system with a negative cosmological constant in arbitrary dimensions higher than three, there exists a horizonless charged soliton solution, which we construct explicitly for an arbitrary mass of the scalar in perturbative series in small charge. We find that the stability of the soliton is determined by the validity of the AdS weak gravity conjecture.
The existence of a stable soliton might endanger the convexity of the (free) energy as a function of the charge because the phase transition between the soliton and the extremal Reissner-Nordstrom black hole would be discontinuous.
We, however, argue that the existence of the hairy black hole solution circumvents the violation of convexity. 
The thermodynamic properties of the hairy black hole show that the phase transition becomes continuous irrespective of whether the AdS weak gravity conjecture holds.
When it holds, the phase transition occurs between the soliton and the hairy black hole, and when it is violated, the phase transition occurs between the extremal Reissner-Nordstrom black hole and the hairy black hole.
\end{abstract}

\thispagestyle{empty} 

\setcounter{page}{0}

\newpage
\section{Introduction}

Convexity of the (free) energy is at the heart of thermodynamic stability \cite{Landau:1980mil}. Still, it is a non-trivial problem whether this is realized in statistical mechanics with a given microscopic Hamiltonian. While it is generically very difficult to verify the convexity, the gravitational system may admit explicit analysis because we can read the thermodynamic properties from the classical solutions, circumventing microscopic statistical mechanical computations.
For example, a Schwarzschild black hole in asymptotically flat spacetime or a small Schwarzschild-AdS black hole shows negative specific heat, which means that the convexity (as a function of temperature) is violated while the microscopic understanding of the violation as well as its implication would be quite challenging.

In this paper, we would like to focus on the convexity of the (lowest) energy as a function of the conserved charge. Again, in generic statistical mechanics, while we expect the convexity of the (free) energy as a function of charge from the thermodynamic stability, it is a highly non-trivial task to verify it in interacting systems. 
The gravitational benchmark would be an extremal Reissner-Nordstrom black hole, which does show convexity.

The recent developments in theoretical physics, however, seem to suggest that the (extremal) Reissner-Nordstrom black hole is not the lowest energy state with a given charge. The weak gravity conjecture \cite{Arkani-Hamed:2006emk}, for example, states that it should be unstable in any consistent theory of quantum gravity (see e.g. \cite{Harlow:2022ich} for review). Indeed, with (light) charged matter, we observe its instability caused by various mechanisms such as classical instability from superradiance or quantum tunneling (e.g. Schwinger effect). Then the task of verifying the convexity requires a study of the onset of such instability and becomes more involved, which we will undertake in this paper.

In terms of the AdS/CFT correspondence, the convexity of the global energy in the AdS space-time as a function of the charge is related to the charge convexity conjecture of the conformal dimensions in conformal field theories. This was first proposed in \cite{Aharony:2021mpc} and then checked in many examples \cite{Antipin:2021rsh}\cite{Moser:2021bes}\cite{Palti:2022unw}\cite{Orlando:2023ljh}\cite{Aharony:2023ike}.\footnote{The convexity and the superadditivity are closely related but they can differ. Strictly speaking, in the original paper \cite{Aharony:2021mpc}, the charge superadditivity is conjectured. See Appendix~\ref{sec:conv_vs_spad} for the connection. See \cite{Antipin:2021rsh} as well.} A counter-example in three-dimensional conformal field theories was found in \cite{Sharon:2023drx}, whose significance should be understood better. A closely related subject is the interpretation of the weak gravity conjecture in terms of conformal field theories \cite{Nakayama:2015hga}. We do have a general picture in many examples, but the precise bound has not been established.

\begin{figure}[tb]
    \centering
    \includegraphics[keepaspectratio, scale=1.0]{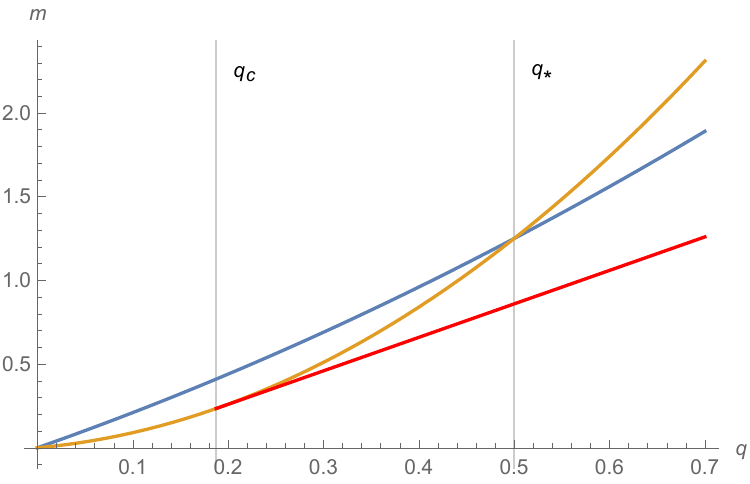}
    \caption{A schematic plot for the mass (or global energy) of the (extremal) Reissner-Nordstrom black hole (blue), charged soliton (orange), and the hairy black hole (red) in $d=3$. Here, the AdS weak gravity conjecture is satisfied. See section~\ref{sec:PT} for details.}
    \label{fig:hairy}
\end{figure}

In this paper, we investigate solutions of the Einstein-Maxwell-charged scalar system with an arbitrary mass for the scalar in asymptotically AdS spacetime. We find there exists a horizonless charged soliton solution, which may have a lower global energy than the extremal Reissner-Nordstrom black hole. We constructed them in perturbative series in small charge, and they are generalizations of massless cases in five dimensions \cite{Basu:2010uz}\cite{Bhattacharyya:2010yg}\cite{Dias:2011tj} to arbitrary dimensions with arbitrary mass.\footnote{A direct motivation of our paper comes from Figures 2 and 3 in \cite{Dias:2011tj}, which show the apparent violation of the charge convexity. Their figures were ``schematic" as they say, and if we closely follow their computations, their results do not violate the convexity. Compare them with Fig.~\ref{fig:hairy}.} See also charged solitons in four dimensions with and without scalar masses \cite{Maeda:2010hf}\cite{Gentle:2011kv}\cite{Dias:2016pma}. We first point out that the existence of such a charged soliton might endanger the convexity of the lowest energy as a function of the charge because the phase transition between the soliton and the extremal Reissner-Nordstrom black hole would be discontinuous.\footnote{In this paper, we consider the canonical ensemble at zero temperature. Meanwhile, the phase transition between the (finite temperature) charged AdS black hole and soliton is often discussed in the grand canonical ensemble \cite{Basu:2016mol} through the grand potential. The convexity of the (canonical) free energy discussed in this paper is equivalent to the concavity of the grand potential due to the thermodynamic relation \cite{Landau:1980mil}.}
We, however, argue that the convexity is restored thanks to the existence of the hairy black hole solution. In particular, we show that the existence of the hairy black hole makes the phase transition continuous.

When the AdS weak gravity conjecture is satisfied by the scalar field, the soliton is lighter than the extremal Reissner-Nordstrom black hole of the same charge in the small charge limit. If we increase the charge, the charged soliton becomes a hairy black hole, and the latter will be the lowest energy state. When the AdS weak gravity conjecture is violated, the soliton is heavier than the extremal Reissner-Nordstrom black hole in the small charge limit. If we increase the charge, the extremal Reissner-Nordstrom black hole becomes unstable to form the hairy black hole, which will be the lowest energy state. In both cases, the phase transition will be continuous and the convexity will be restored. A schematic plot for the case when the AdS weak gravity conjecture is satisfied is given in Fig.~\ref{fig:hairy}.

The organization of the paper is as follows. In section~\ref{sec:RN_and_soliton}, we construct the charged soliton solution in perturbative series in small charge and compare the thermodynamic properties with those of the extremal Reissner-Nordstrom black hole. In section~\ref{sec:hairyBH}, we study the thermodynamic properties of a hairy black hole. In section~\ref{sec:PT}, we study the phase transition between these solutions to see if the convexity is violated or not. Section~\ref{sec:discussion} is devoted to discussion. Appendices contain details of calculations.

\section{Reissner-Nordstrom black hole and charged soliton}
\label{sec:RN_and_soliton}

We study solutions of the Einstein-Maxwell-charged scalar theory described by the action
\begin{align}
S = \int d^{d+1}x \sqrt{-g} \left( \frac{1}{16 \pi G_N} (R - 2 \Lambda) - \frac{1}{4}F_{\mu\nu} F^{\mu\nu} - |D_\mu \phi|^2 - m_\phi^2 |\phi|^2\right) \ ,
\label{action}
\end{align}
where $\Lambda = -d(d-1)/2$ is the negative cosmological constant in units in which the AdS radius is unity, and the gauge covariant derivative is defined as $D_\mu \phi = \partial_\mu \phi - i e A_\mu \phi$ with $e$ being the $U(1)$ coupling constant.
For convenience, we will parameterize the mass of the charged scalar by the conformal dimension $\Delta$ as $m_\phi^2 = \Delta(\Delta-d)$. In this paper, we assume $d \ge 3$.

Without the scalar hair $(\phi=0)$, the supposedly lowest energy solution with charge $Q$ is the Reissner-Nordstrom-AdS black hole solution. In the conventional radial coordinate, the Reissner-Nordstrom-AdS black hole has the metric and gauge field:
\begin{align}
ds^2 &= -f(r) dt^2 + \frac{dr^2}{f(r)} + r^2 d \Omega_{d-1} \ , \cr
A_\mu dx^\mu &= \left( \mu - \frac{q}{r^{d-2}} \right) dt \ , \cr
f(r) &= r^2 + 1 - \frac{m}{r^{d-2}} + \frac{d-2}{d-1}\frac{\kappa^2 q^2}{r^{2(d-2)}} \ ,
\label{eq_RN_solution}
\end{align}
where we use the notation $\kappa^2 = 8 \pi G_N$. The event horizon of the black hole is located at $r=r_h$ given by the largest root of $f(r_h)=0$. We will fix the $U(1)$ gauge by demanding the gauge field satisfies $A_t(r_h)=0$, which specifies that $\mu$ is the chemical potential of the $U(1)$. From these conditions, we obtain
\begin{align}
m &= r_h^{d-2} \left( r_h^2 + 1 + \frac{d-2}{d-1}\frac{\kappa^2 q^2}{r_h^{2(d-2)}} \right) \ , \cr
\mu &= \frac{q}{r_h^{d-2}} \ .
\label{eq_m_mu_RN}
\end{align}

The thermodynamic properties of the Reissner-Nordstrom-AdS black hole can be obtained from the above solution in a standard manner. The global energy and charge are related to the parameters in the solution as
\begin{align}
E &= \frac{d-1}{16 \pi G_N} \omega_{d-1}m \ , \cr
Q &= (d-2) \omega_{d-1} q \ ,
\label{eq_EQ_by_mq}
\end{align}
where $\omega_{d-1}$ denotes the volume of a unit $(d-1)$-sphere, $\omega_{d-1} = 2 \pi^{\frac{d}{2}}/\Gamma(\frac{d}{2})$. The chemical potential can then be written as a function of the charge as
\begin{align}
\mu = \frac{Q}{(d-2)r_h^{d-2}\omega_{d-1}} \ .
\end{align}
The temperature and the entropy are given by
\begin{align}
T &= \frac{1}{4 \pi} f'(r_h) \ , \cr
S &= \frac{r_h^{d-1} \omega_{d-1}}{4 G_N} \ .
\end{align}
These thermodynamic quantities satisfy the first law of thermodynamics $dE = T dS + \mu dQ$.

In the following, we focus on the zero temperature limit, where
the Reissner-Nordstrom black hole becomes extremal. 
Note that in the extremal limit, we have $\mu = \frac{\partial E}{\partial Q}$ because the extremal black hole has $T=0$ so $F=E-TS =E$, and $\mu = \frac{\partial F}{\partial Q} = \frac{\partial E}{\partial Q}$.
Then, we want to express the energy $E$ as a function of charge $Q$ explicitly. We can obtain the closed formulae valid for any charge in some particular dimensions, say $d=3,4$ and $5$.\footnote{Eq.~\eqref{eq_extremal_rh} is a quadratic, cubic, or quartic equation of $r_h^2$ in $d=3,4$, or $5$, respectively, and can be solved for $r_h^2$, but it cannot be algebraically solved in general in $d \ge 6$, where the equation is quintic or of higher degrees.} For example, in $d=3$, we find  (see e.g. \cite{Loukas:2018zjh}\cite{Nakayama:2020dle})
\begin{align}
E_{d=3} = \frac{4 \pi \sqrt{6}}{9 \kappa^2} \sqrt{-1+\sqrt{1+6 \kappa^2 q^2} + 6 \kappa^2 q^2 \left(3+\sqrt{1+6 \kappa^2 q^2} \right)} \ ,
\end{align}
where $q$ is related to $Q$ as in \eqref{eq_EQ_by_mq}. In this paper, instead, we will only use the small charge expansion.
The explicit form of the energy and chemical potential in the small charge expansion is given in general dimensions by (see Appendix~\ref{sec:small_charge_RN_derivation} for calculations)
\begin{align}
E &= \frac{1}{\kappa}\sqrt{\frac{d-1}{d-2}} Q \left( 1 + \frac{1}{2} \widetilde{Q}^{\frac{2}{d-2}} - \frac{d^2}{8(d-2)^2} \widetilde{Q}^{\frac{4}{d-2}} + \frac{d^4}{16(d-2)^4} \widetilde{Q}^{\frac{6}{d-2}} + \cdots \right), \cr
\mu &=  \frac{1}{\kappa}\sqrt{\frac{d-1}{d-2}} \left( 1 + \frac{d}{2(d-2)} \widetilde{Q}^{\frac{2}{d-2}} - \frac{d^2(d+2)}{8(d-2)^3} \widetilde{Q}^{\frac{4}{d-2}} + \frac{d^4(d+4)}{16(d-2)^5} \widetilde{Q}^{\frac{6}{d-2}} + \cdots \right),
\label{eq_RN_Emu_Qseries}
\end{align}
where
\begin{align}
\widetilde{Q} \equiv \frac{\kappa Q}{\omega_{d-1}\sqrt{(d-1)(d-2)}} = \kappa q \sqrt{\frac{d-2}{d-1}} \ .
\label{eq_widetildeQ}
\end{align}

At this point, we can verify that the global energy of the extremal Reissner-Nordstrom black hole is a convex function of the charge. Since $E(0)=0$, this implies that it is superadditive (see Appendix~\ref{sec:conv_vs_spad}). It is a non-trivial prediction of the black hole solution that the expansion parameter is $Q^{\frac{2}{d-2}}$, which is dimension specific (rather than $Q$), and it should be contrasted with the charged soliton solution whose expansion parameter is $Q$ in any dimensions as we will see later.

The slope of the energy as a function of the charge sets the weak gravity conjecture bound. In this paper, we say that the weak gravity conjecture is satisfied when there is a state whose energy-charge ratio is smaller than that of the extremal Reissner-Nordstrom black hole in the small charge limit:
\begin{align}
\frac{E (Q)}{Q} \le  \frac{1}{\kappa}\sqrt{\frac{d-1}{d-2}} \ .
\end{align}
Alternatively, we say that the weak gravity conjecture is violated when there is no state whose energy-charge ratio is smaller than that of the extremal Reissner-Nordstrom black hole in the small charge limit: all the states satisfy
\begin{align}
\frac{E (Q)}{Q} \ge  \frac{1}{\kappa}\sqrt{\frac{d-1}{d-2}} \ .
\end{align}

In the literature, there are some different notions of AdS weak gravity conjecture than that used here (see e.g. \cite{Nakayama:2015hga}\cite{Giombi:2017mxl}\cite{Urbano:2018kax}\cite{Montero:2018fns}\cite{Alday:2019qrf}\cite{Cho:2023koe} in relation to CFTs). At least for the (non-interacting) charged scalar matter, we will see that this bound is relevant in our analysis of the hairy black hole. We also note that the bound we consdier is the same as the bound used in 
\cite{Crisford:2017gsb} to avoid the formulation of the naked singularity in the Einstein-Maxwell-charged scalar system in AdS (in $d=3$ studied there). We will discuss this point further below when we introduce the critical coupling $e_c$.

In addition to the extremal Reissner-Nordstrom black hole, in this theory, there exists a charged soliton solution, that has non-zero scalar condensate. When the weak gravity conjecture holds, we expect that the soliton solution, at least in the small charge limit, has a lower energy than the extremal Reissner-Nordstrom black hole.

We have explicitly constructed the horizonless soliton solution within a perturbative expansion with respect to the scalar condensate (or equivalently charge). See Appendix~\ref{sec:soliton_derivation} for details. They are reliable in the small charge limit, which we will focus on. Some examples of the solutions in various dimensions with various scalar masses can be found in Appendix~\ref{sec:soliton_derivation}. Here we present the final result of the energy and the chemical potential of the soliton as functions of the charge up to the second order in small charge expansion:
\begin{align}
E &=(d-2)\omega_{d-1} \left[ \frac{\Delta}{e} q + \frac{\Gamma(\Delta)^2 \, \Gamma(2\Delta+1-d/2)}{2 \, \Gamma(d/2) \, \Gamma(2\Delta) \, \Gamma(\Delta+1-d/2)^2} \left( 1 - \frac{(d-2)\Delta^2(\Delta-d/2)}{(d-1)(\Delta-d/4)} \frac{\kappa^2}{e^2} \right) q^2 \right] \cr
&= Q \left[ \frac{\Delta}{e} + \frac{\Gamma(\Delta)^2 \, \Gamma(2\Delta+1-d/2)}{2 \, \Gamma(d/2) \, \Gamma(2\Delta) \, \Gamma(\Delta+1-d/2)^2} \left( 1 - \frac{(d-2)\Delta^2(\Delta-d/2)}{(d-1)(\Delta-d/4)} \frac{\kappa^2}{e^2} \right) q \right] \ , \cr
\mu &= \frac{\Delta}{e} + \frac{\Gamma(\Delta)^2 \, \Gamma(2\Delta+1-d/2)}{\Gamma(d/2) \, \Gamma(2\Delta) \, \Gamma(\Delta+1-d/2)^2}\left( 1 - \frac{(d-2)\Delta^2(\Delta-d/2)}{(d-1)(\Delta-d/4)} \frac{\kappa^2}{e^2} \right) q \ ,
\label{eq_soliton_Emu_Qseries}
\end{align}
where $q$ is related to $Q$ as in \eqref{eq_EQ_by_mq}. We see that $\mu = \frac{\partial E}{\partial Q}$, which implies $TS=0$ because $E=F$. There is no horizon, so the gravitational entropy is zero. In the Euclidean continuation of the horizonless soliton, the periodicity of the compactified time direction is arbitrary, so is the temperature. In this paper, we consider zero temperature.

Let us introduce the notion of the critical coupling $e_c$. When the charge $Q$ is small, the energy of the extremal Reissner-Nordstrom black hole \eqref{eq_RN_Emu_Qseries} and that of the soliton \eqref{eq_soliton_Emu_Qseries} behave as
\begin{align}
E_{RN} \simeq \frac{1}{\kappa}\sqrt{\frac{d-1}{d-2}} Q \ , \qquad
E_{sol} \simeq \frac{\Delta}{e} Q \ .
\end{align}
We identify the critical coupling $e_c$ as the value of the coupling that equates these energies,
\begin{align}
e_c = \Delta \kappa \sqrt{\frac{d-2}{d-1}} \ .
\label{eq_ec}
\end{align}
This critical coupling is exactly the lower bound of the gauge coupling constant to preserve cosmic censorship discussed in \cite{Crisford:2017gsb} for $d=3$.\footnote{In \cite{Crisford:2017gsb}, the lower bound of the scalar field charge (i.e. $U(1)$ coupling constant) $q_W$ is given in their normalization as $q_W^{(\text{theirs})}=\Delta$, where the AdS radius is set unity, but our normalization of the fields and coupling constant is different from theirs. Rescaling the gauge field and gauge coupling constant in \cite{Crisford:2017gsb} so that they match ours, the lower bound reads $q_W^{(\text{ours})}=\Delta \kappa/\sqrt{2}$, which is nothing but \eqref{eq_ec} for $d=3$: $q_W^{(\text{ours})} = e_c|_{d=3}$.} We expect \eqref{eq_ec} gives the generalization of the bound of \cite{Crisford:2017gsb} to general dimensions $d \ge 3$.

The critical coupling connects the soliton with the weak gravity conjecture mentioned above. Given $\Delta$ and small $Q$, the small charged soliton has a smaller energy than the extremal Reissner-Nordstrom black hole when $e>e_c$. This is equivalent to saying that the weak gravity conjecture is satisfied by the charged soliton. We can also say that gravity is weaker than the electric force and a horizonless charged soliton can be formed without gravitationally collapsing to a black hole. Alternatively, the small charged soliton has a larger energy than the extremal Reissner-Nordstrom black hole when $e<e_c$. This is equivalent to saying that the weak gravity conjecture does not hold because the electric force is weaker than gravity.

Another point we would like to discuss about the properties of the charged soliton is the convexity of the energy. The energy of the soliton is convex $\frac{d^2 E}{d Q^2} > 0$ when the coupling constant satisfies
\begin{align}
e^2 > \frac{(d-2)\Delta^2(\Delta-d/2)}{(d-1)(\Delta-d/4)} \kappa^2 = \frac{\Delta-d/2}{\Delta-d/4} e_c^2 \ .
\label{eq_soliton_convexity_coupling}
\end{align}
The right-hand side is not positive for $ \Delta \le d/2$ in $d \geq 4$ and $3/4 < \Delta \le 3/2$ in $d=3$.\footnote{Recall that the range that $\Delta$ can take is $\Delta \ge d/2-1$ (unitarity bound).} This means that, when $\Delta$ is in these ranges, the energy of the soliton is convex for any coupling constant $e$.
Otherwise, the right-hand side is positive. Accordingly, if the coupling constant is too small, the energy of the soliton is not convex. However, such a coupling is smaller than the critical coupling \eqref{eq_ec} (for $\Delta > d/2$) below which the soliton is not the configuration with the smallest energy, and the convexity is irrelevant. If we further believe in the weak gravity conjecture, this parameter region does not arise.

We here observe that in $d=3, 4$ the inequality \eqref{eq_soliton_convexity_coupling} obtained from solving Einstein's equation is equivalent to demanding positivity of the binding energy $\gamma_{\text{grav}} + \gamma_{\text{photon}} > 0$ from the perturbative formula in \cite{Fitzpatrick:2011hh} ($d=4$) and \cite{Andriolo:2022hax} ($d=3$). See also eq.~(2.5) of \cite{Aharony:2021mpc} and discussions there. We expect a similar comparison can be done in other dimensions.

The only subtle case is $1/2 < \Delta < 3/4$ for $d=3$, where the right-hand side of the bound \eqref{eq_soliton_convexity_coupling} is larger than the critical coupling, i.e. $e^2 >4e_c^2$. This means that the convexity can be violated near $e\sim e_c$. Our formula implies that the attractive force from the graviton exchange (i.e. $\gamma_{\text{grav}}$ in \cite{Andriolo:2022hax}) becomes infinite at $\Delta = 3/4$ and continues to be large down to the unitarity bound $\Delta = 1/2$. We might want to say that some extended notion of weak gravity conjecture is violated in this case even though $e >e_c$. If the soliton is stable, which we do not know, it may violate the charge convexity. We, however, note that our formula for $1/2 < \Delta < 3/4$ is obtained from the analytic continuation of $\Delta > 3/4$ and we have not constructed an explicit solution in this range, which may be singular.
To avoid the subtlety, we assume $\Delta > 3/4$ for $d=3$ hereafter.

\section{Hairy black hole}
\label{sec:hairyBH}
As we have seen, if the weak gravity conjecture is violated, even with the existence of the soliton solution, the extremal Reissner-Nordstrom black hole solution has a lower energy in the small charge limit. What happens if we increase the charge? Does it show a phase transition to the charged soliton? Or, does it continue to be the lowest energy state?
In the following, we will argue that at a certain critical charge, the extremal Reissner-Nordstrom black hole forms a scalar hair. The formation of the hair is due to superradiant instability of the Reissner-Nordstrom-AdS black hole \cite{Dias:2016pma}\cite{Hawking:1999dp}\cite{Uchikata:2011zz}\cite{Green:2015kur}. Similarly, if the weak gravity conjecture holds, the lowest energy solution with a given charge is the soliton (in the small charge limit), but if we increase the charge further (as well as the energy), it collapses into a hairy black hole. In this section, we will predict the thermodynamic properties of the hairy black hole in the Einstein-Maxwell-charged scalar system, and in the next section, we will study the nature of the phase transition.

We have already seen that both the horizonless charged soliton solution and the Reissner-Nordstrom black hole solution are compatible with the thermodynamic relation $\mu= \frac{\partial E}{\partial Q}$ at zero temperature. It is natural to expect that the hairy black hole solution also obeys the thermodynamic relations. This observation enables us to use the thermodynamic equilibrium argument to predict the property of the hairy black hole as well. As discussed in \cite{Dias:2011tj}, when the charge is small, we will regard the extremal hairy black hole as a thermodynamic mixture of the extremal Reissner-Nordstrom black hole and the horizonless charged soliton. We express the charge and the energy of the hairy black hole as a non-interacting sum as 
\begin{align}
Q_{hairy} &= a Q_{RN} + b Q_{sol} \ , \cr
E_{hairy} &= a E_{RN}(Q_{RN}) + bE_{sol}(Q_{sol}) \ ,
\label{equilibrium}
\end{align}
where the mixing parameters $a,b \ge0$ are determined by the thermodynamic equilibrium condition
\begin{align}
\mu_{RN} (Q_{RN}) = \mu_{sol} (Q_{sol}) \ . 
\end{align}
The thermodynamic equilibrium condition also demands that the temperature must be equal, and our assumption that the charged soliton has zero temperature makes it equilibrium with the extremal Reissner-Nordstrom black hole.

This simple picture gives a determination of the critical charge $Q_c$ above which we will see that the hairy black hole becomes the lowest energy state. When $e<e_c$
(i.e. weak gravity conjecture is violated), it is determined by equating the chemical potential of the Reissner-Nordstrom black hole at $Q_{RN}=Q_c$ and that of the soliton at $Q_{sol}=0$: $\mu_{RN} (Q_c) = \mu_{sol} (0)$.
In the same way, when $e>e_c$ (i.e. weak gravity conjecture holds), it is determined by equating the chemical potential of the soliton at $Q_{sol}=Q_c$ and that of the Reissner-Nordstrom black hole at $Q_{RN}=0$:  $\mu_{RN} (0) = \mu_{sol} (Q_c)$. 

The critical charge $Q_c$ can be analytically expressed in arbitrary dimensions.\footnote{For $d=3$, we assume $\Delta > \frac{3}{4}$. See discussions after \eqref{eq_soliton_convexity_coupling}.}
For $e <e_c$, let us take $e^2 = e_c^2(1 - \theta) < e_c^2$, where $\theta>0$ is a small parameter.
With this coupling, the chemical potential of the soliton at $Q_{sol}=0$ is given by
\begin{align}
\mu_{sol}(0) = \frac{\Delta}{e} = \frac{1}{\kappa} \sqrt{\frac{d-1}{d-2}} \left( 1 + \frac{\theta}{2} + \cdots \right) \ .
\end{align}
Comparing this with $\mu_{RN}(Q_c)$ in \eqref{eq_RN_Emu_Qseries} to the subleading terms, we obtain
\begin{align}
Q_c = \frac{\omega_{d-1}\sqrt{(d-1)(d-2)^{d-1}}}{\kappa} \left( \frac{\theta}{d} \right)^{\frac{d-2}{2}} \ .
\label{Qc_e_less_ec}
\end{align}
For $e>e_c$, we take $e^2 = e_c^2(1+\theta) > e_c^2$, where $\theta>0$. The charge of the soliton that balances with this shift of the coupling can be parametrized as $Q_c = q_{sol} = b_c \theta$. Then, the series expansion of \eqref{eq_soliton_Emu_Qseries} in $\theta$ becomes
\begin{align}
\mu_{sol} (Q_c) = \frac{1}{\kappa} \sqrt{\frac{d-1}{d-2}} + \left( \frac{b_c d \, \Gamma(\Delta)^2 \, \Gamma(2\Delta+1-d/2)}{(4\Delta-d)\Gamma(d/2) \, \Gamma(2\Delta) \, \Gamma(\Delta+1-d/2)^2} - \frac{1}{2\kappa} \sqrt{\frac{d-1}{d-2}} \right) \theta + \cdots \ .
\label{mu_soliton_e_gtr_ec}
\end{align}
The leading order term is the same as $\mu_{RN}(0)$ from eq.~\eqref{eq_RN_Emu_Qseries}. Then, from $\mu_{RN} (0) = \mu_{sol} (Q_c)$, the critical charge is given by the vanishing of the subleading term in \eqref{mu_soliton_e_gtr_ec} as
\begin{align}
Q_c = \frac{4\Delta-d}{2 \kappa d} \omega_{d-1} \sqrt{(d-1)(d-2)} \frac{\Gamma(d/2) \, \Gamma(2\Delta) \, \Gamma(\Delta+1-d/2)^2}{\Gamma(\Delta)^2 \, \Gamma(2\Delta+1-d/2)} \theta \ .
\end{align}

Now we have determined the critical charge, let us study the energy of the hairy black hole above $Q_c$. We solve the equilibrium condition \eqref{equilibrium} with suitable ansatz in the small charge limit.
The results are summarized below for $d=3,4,5$ and $Q>Q_c$. We will also give comments on the hairy black holes in $d \ge 6$. For $Q<Q_c$, we find that the energy of the hairy black hole is larger than that of the soliton or extremal Reissner-Nordstrom black hole (and one of the mixing parameters in \eqref{equilibrium} becomes negative), so below we will not discuss the hairy black holes with $Q<Q_c$. For derivation, see Appendix~\ref{sec:hairy_derivation}.

\subsection{Hairy black hole in $e < e_c$}

When the coupling constant is smaller than the critical value as $e^2 = e_c^2 (1 - \theta)$, we can construct a hairy black hole with $Q \sim \theta^{(d-2)/2}$. 

\paragraph{AdS$_4$ $(d=3)$}
The critical coupling is
\begin{align}
Q_c = \omega_2 q_c \ , \qquad
q_c = \frac{\sqrt{6}}{3 \kappa} \theta^{1/2} \ .
\end{align}
The energy of the hairy black hole in $Q > Q_c$ is
\begin{align}
E = \frac{\omega_2}{\kappa^2} m_{hairy} \ , \qquad
m_{hairy} = m_{RN} - \frac{3 \times 2^{2 \Delta - 5} \, \Gamma(\Delta-1/2)^2 \, \Gamma(\Delta+1/2)}{\Gamma(\Delta) \, \Gamma(2\Delta-3/2)} \kappa^4 (q^2 - q_c^2)^2 \ ,
\end{align}
where
\begin{align}
m_{RN} = \sqrt{2} \kappa q + \frac{\kappa^3 q^3}{2\sqrt{2}} + O(q^5) \ .
\label{eq_mRN_e_less_ec_d3}
\end{align}
The energy satisfies $m_{hairy} < m_{RN}$ if $\Delta>3/4$. The energy of the hairy black hole is a convex function of the charge as we can see at $O(q^3)$ of the Reissner-Nordstrom black hole \eqref{eq_mRN_e_less_ec_d3}.

\paragraph{AdS$_5$ $(d=4)$}
The critical coupling is
\begin{align}
Q_c = 2 \omega_3 q_c \ , \qquad
q_c = \frac{\sqrt{6}}{4 \kappa} \theta \ .
\end{align}
The energy of the hairy black hole in $Q > Q_c$ is
\begin{align}
E = \frac{3 \omega_3}{2\kappa^2} m_{hairy} \ , \qquad
m_{hairy} = m_{RN} - \frac{2 (2 \Delta - 1)}{3 (3\Delta - 2)} \kappa^2 (q - q_c)^2 \ ,
\end{align}
where
\begin{align}
m_{RN} = \frac{2\sqrt{6}}{3} \kappa q + \frac{2}{3} \kappa^2 q^2 \ .
\end{align}
For $\Delta>1$ ($=\frac{d}{2}-1$ with $d=4$), we have $\frac{2}{3} < \frac{2 \Delta - 1}{3\Delta - 2} < 1$. Hence, we find that the energy of the hairy black hole satisfies $m_{hairy} < m_{RN}$ and also is a convex function of the charge.

\paragraph{AdS$_6$ $(d=5)$}
The critical coupling is
\begin{align}
Q_c = 3 \omega_4 q_c \ , \qquad
q_c = \frac{6}{5 \sqrt{5} \kappa} \theta^{3/2} \ .
\end{align}
The energy of the hairy black hole in $Q > Q_c$ is
\begin{align}
E = \frac{2 \omega_4}{\kappa^2} m_{hairy} \ , \qquad
m_{hairy} &= \sqrt{3} \kappa q + \left( \frac{\sqrt{3}}{2} \kappa q \theta - \frac{6 \sqrt{15}}{125} \theta^{5/2} \right) \cr
&= m_{RN} - \left( \frac{3^{5/6} \kappa^{5/3} q^{5/3}}{2^{5/3}} - \frac{\sqrt{3}}{2} \kappa q \theta + \frac{6 \sqrt{15}}{125} \theta^{5/2} \right) \ ,
\label{eq_ads6_e_less_ec}
\end{align}
where
\begin{align}
m_{RN} = \sqrt{3} \kappa q + \frac{3^{5/6} \kappa^{5/3} q^{5/3}}{2^{5/3}} \ .
\end{align}
One can check that $m_{hairy} < m_{RN}$ and $\frac{d m_{hairy}}{d q}|_{q=q_c} = \frac{d m_{RN}}{d q}|_{q=q_c}$ by expanding \eqref{eq_ads6_e_less_ec} around $q=q_c$, where $|q-q_c| \ll \theta^{3/2}$, as
\begin{align}
m_{hairy} = m_{RN} - \frac{5 \sqrt{15}}{36 \sqrt{\theta}} \kappa^2 (q-q_c)^2 + O \left( (q-q_c)^3 \right) \ .
\end{align}
The result \eqref{eq_ads6_e_less_ec} given up to $O(q)$ is not sufficient to discuss the convexity of $m_{hairy}$, so we also calculate the next order $O(q^2)$ term and find the convexity as
\begin{align}
\frac{d^2 m_{hairy}}{d q^2} = \frac{5 \kappa^2 \, \Gamma(\Delta) \, \Gamma(2\Delta-5/2)}{2^{2 \Delta-1} \, \Gamma(\Delta+1/2) \, \Gamma(\Delta-3/2)^2} >0 \ .
\end{align}

\subsection{Hairy black hole in $e > e_c$}

When the coupling constant is larger than the critical value as $e^2 = e_c^2 (1 + \theta)$, we can construct a hairy black hole with $Q \sim \theta$.

\paragraph{AdS$_4$ $(d=3)$}
The critical coupling is
\begin{align}
Q_c = \omega_2 q_c \ , \qquad
q_c = \frac{2^{2 \Delta - 3/2} \, \Gamma(\Delta-1/2)^2 \, \Gamma(\Delta+1/2)}{3 \kappa \, \Gamma(\Delta) \, \Gamma(2\Delta-3/2)} \theta \ .
\end{align}
The energy of the hairy black hole in $Q > Q_c$ is
\begin{align}
E = \frac{\omega_2}{\kappa^2} m_{hairy} \ , \qquad
m_{hairy} &= \sqrt{2} \kappa q - \frac{ 2^{2\Delta-3} \, \Gamma(\Delta-1/2)^2 \, \Gamma(\Delta+1/2)}{3 \, \Gamma(\Delta) \, \Gamma(2\Delta-3/2)} \theta^2 \cr
&= m_{sol} - \frac{3 \, \Gamma(\Delta) \, \Gamma(2\Delta-3/2)}{2^{2 \Delta} \, \Gamma(\Delta-1/2)^2 \, \Gamma(\Delta+1/2)} \kappa^2 (q - q_c)^2 \ ,
\end{align}
where
\begin{align}
m_{sol} = \sqrt{2} \kappa q + \left(\frac{3 \, \Gamma(\Delta) \, \Gamma(2\Delta-3/2)}{2^{2 \Delta} \, \Gamma(\Delta-1/2)^2 \, \Gamma(\Delta+1/2)} \kappa^2 q^2 - \frac{\kappa q \theta}{\sqrt{2}} \right) \ .
\end{align}
The energy satisfies $m_{hairy} < m_{sol}$ if $\Delta>3/4$. Because the coefficient of $q^2$ in $m_{hairy}$ vanishes, we need to go to the next order to check the convexity. From the calculations including $O(\theta^3)$ terms, we find
\begin{align}
\frac{d^2 m_{hairy}}{d q^2} = \frac{3\sqrt{2}}{2} \kappa^3 (q-q_c) \ ,
\end{align}
so the convexity is satisfied in $Q>Q_c$.

\paragraph{AdS$_5$ $(d=4)$}
The critical coupling is
\begin{align}
Q_c = 2 \omega_3 q_c \ , \qquad
q_c = \frac{\sqrt{6} (2 \Delta - 1)}{4 (\Delta - 1) \kappa} \theta \ .
\end{align}
The energy of the hairy black hole in $Q > Q_c$ is
\begin{align}
E = \frac{3 \omega_3}{2\kappa^2} m_{hairy} \ , \qquad
m_{hairy} = m_{sol} - \frac{2 (\Delta - 1)^2}{3 (2\Delta - 1)(3\Delta - 2)} \kappa^2 (q - q_c)^2 \ ,
\end{align}
where
\begin{align}
m_{sol} = \frac{2\sqrt{6}}{3} \kappa q + \left(\frac{2(\Delta-1)}{3 (2 \Delta - 1)} \kappa^2 q^2 - \frac{\sqrt{6}}{3} \kappa q \theta \right) \ .
\end{align}
The energy satisfies $m_{hairy} < m_{sol}$. Because $1-\frac{\Delta - 1}{3\Delta - 2}$ is positive in $\Delta>1$, the convexity is satisfied.

\paragraph{AdS$_6$ $(d=5)$}
The critical coupling is
\begin{align}
Q_c = 3 \omega_4 q_c \ , \qquad
q_c = \frac{2^{2 \Delta-2} \sqrt{3} \, \Gamma(\Delta-3/2)^2 \, \Gamma(\Delta+1/2)}{5 \kappa \, \Gamma(\Delta) \, \Gamma(2\Delta-5/2)} \theta \ .
\end{align}
The energy of the hairy black hole in $Q > Q_c$ is
\begin{align}
E = \frac{2 \omega_4}{\kappa^2} m_{hairy} \ , \qquad
m_{hairy} = m_{sol} - \frac{3^{1/4} \, \Gamma(\Delta)^{5/2} \, \Gamma(2\Delta-5/2)^{5/2}}{2^{5 \Delta - 6} \, \Gamma(\Delta-3/2)^5 \, \Gamma(\Delta+1/2)^{5/2}} \kappa^{5/2} (q-q_c)^{5/2} \ , \label{hairyd5}
\end{align}
where
\begin{align}
m_{sol} = \sqrt{3} \kappa q + \left(\frac{5 \, \Gamma(\Delta) \, \Gamma(2\Delta-5/2)}{2^{2 \Delta} \, \Gamma(\Delta-3/2)^2 \, \Gamma(\Delta+1/2)} \kappa^2 q^2 - \frac{\sqrt{3}}{2} \kappa q \theta \right) \ .
\label{eq_ads6_msol_ref}
\end{align}
We can see that $m_{hairy} < m_{sol}$. The convexity of $m_{hairy}$ is satisfied because the energy of the soliton \eqref{eq_ads6_msol_ref} is already convex and it dominates over the added term in \eqref{hairyd5}.

Let us conclude this section with comments on hairy black holes in $d \ge 6$. When $e<e_c$ (i.e. the weak gravity conjecture is violated), the similar construction outlined in Appendix~\ref{sec:hairy_derivation} leads to the expression $m_{hairy}(q) = m_{RN}(q) + C_{RN}(q-q_c)^{2} + O((q-q_c)^3) $ around $q=q_c$. The coefficient $C_{RN}$ is such that $q^{d/d-2}$ term in the small $q$ expansion of $m_{hairy}(q)$ vanishes and $m_{hairy}(q)$ is an integer power series in $q$. The above form of $m_{hairy}(q)$ implies that the phase transition is continuous between the Reissner-Nordstrom black hole and the hairy black hole (i.e. the first derivatives of $m$ with respect to $q$ are the same at $q=q_c$). The convexity of $m_{hairy}(q)$ can be checked from the explicit computation of the $O(q^2)$ term.

When $e>e_c$ (i.e. the weak gravity conjecture holds), the similar ansatz in Appendix~\ref{sec:hairy_derivation} instead leads to the expression $m_{hairy}(q) = m_{sol}(q) + C_{sol}(q-q_c)^{\frac{d-2}{2}}$ around $q=q_c$. To determine the precise numerical constant $C_{sol}$ (independent of $\theta$), we need the explicit form of the soliton to higher orders than we calculated. This is beyond the scope of this paper and we do not attempt. Nevertheless, we can still argue that the energy of the hairy black hole must be convex for $d\ge 6$ because $q^2$ term in $m_{sol}(q)$ is dominant over $C(q-q_c)^{\frac{d-2}{2}}$ in the small charge expansions of $m_{hairy}(q)$ and we know its convexity (as discussed at the end of the previous section).

\section{Phase transition and restoration of convexity }
\label{sec:PT}

Let us now study the would-be phase transition between the Reissner-Nordstrom solution and the soliton solution that we constructed in section~\ref{sec:RN_and_soliton}. Later in this section, within the parameter space where our solution is valid, we will show that this type of phase transition never happens due to the existence of the hairy black hole solution. We, nevertheless, first ask ourselves what would happen if the hairy black hole solution did not exist.

First, consider the case when the weak gravity conjecture is violated (i.e. $e<e_c$). In the small charge limit, the extremal Reissner-Nordstrom black hole has the lowest energy. Depending on the parameter, it may or may not happen that if we increase the charge, the charged soliton solution starts to possess a lower energy than the extremal Reissner-Nordstrom black hole. When it does, generically there is a (would-be) discontinuous phase transition from the Reissner-Nordstrom black hole to the charged soliton solution. It is (generically) discontinuous because the first derivative of the energy with respect to the charge (i.e. chemical potential) at the phase transition point is discontinuous.

Similarly, when the weak gravity conjecture holds (i.e. $e>e_c$), in the small charge limit the charged soliton has the lowest energy. Then there may exist a phase transition to the extremal Reissner-Nordstrom black hole if we increase the charge. If it does, again the phase transition would be discontinuous (see Fig.~\ref{fig:hairy}).

One can compute the would-be phase transition point more explicitly in the small charge limit where our solutions are reliable.
In AdS$_4$, solving $m_{RN}(q_*) = m_{sol}(q_*)$ in the small charge limit (i.e. $|\theta| \ll 1$), we obtain for $e^2=e_c^2(1+\theta)>e_c^2$
\begin{align}
q_* = \frac{2^{2\Delta-1/2} \, \Gamma(\Delta-1/2)^2 \, \Gamma(\Delta+1/2)}{3 \kappa \, \Gamma(\Delta) \, \Gamma(2\Delta-3/2)} \theta \ , 
\end{align}
while there is no $q_*>0$ for $e<e_c$ (we assume $\Delta > 3/4$), so the (would-be) phase transition occurs when $e>e_c$ (i.e. the AdS weak gravity conjecture holds).

In AdS$_5$, a similar calculation gives for $e^2=e_c^2(1-\theta)<e_c^2$
\begin{align}
q_* = \frac{\sqrt{6}(2\Delta-1)}{2 \kappa \Delta} \theta \ ,
\end{align}
while there is no $q_*>0$ for $e>e_c$,
so the (would-be) phase transition occurs when $e<e_c$ (i.e. the AdS weak gravity conjecture does not hold). 

In AdS$_{d+1}$ with $d \ge 5$, we observe that the (would-be) phase transition occurs when $e<e_c$ (i.e. the AdS weak gravity conjecture does not hold). For $e^2=e_c^2(1-\theta)<e_c^2$, we obtain
\begin{align}
q_* = \frac{1}{\kappa} \sqrt{\frac{d-1}{d-2}} \, \theta^{\frac{d-2}{2}} \ .
\end{align}
while there is no $q_*>0$ for $e>e_c$. Note that in all these cases, the scaling of $q_*$ with respect to $\theta$ is the same as that for $q_c$ studied in the previous section.

When the phase transition is discontinuous, the convexity of the energy is violated. The second derivative of the energy as a function of the charge becomes negative (in the delta function sense) at the phase transition, and we can easily find three charges around the phase transition point where the convexity inequality $\lambda E(Q_1) + (1-\lambda) E(Q_2) \ge E(\lambda Q_1 + (1-\lambda) Q_2)$ is violated.

Note, however, that this does {\it not} mean that the discontinuous phase transition causes the violation of the superadditivity. The convex function with $E(0)=0$ is always superadditive (see Appendix~\ref{sec:conv_vs_spad}), but the converse is not necessarily true. Indeed, we can show $E(Q) = \mathrm{min}(E_{RN}(Q), E_{sol}(Q))$ is superadditive. The elementary proof can be found in Appendix~\ref{sec:conv_vs_spad}.

The existence of the hairy black hole solution that we found in the previous section completely changes the story above. This is because it is always the case that the hairy black hole enters the phase diagram before the would-be phase transition discussed above (i.e. $q_c<q_*$). See Fig.~\ref{fig:hairy} for a schematic plot of this situation. In particular, the would-be violation of the convexity of the energy is circumvented as we will see.

The study in section~\ref{sec:hairyBH} shows that above $Q_c$, the hairy black hole is the lowest energy solution of the system with a given charge $Q$.\footnote{Logically speaking, we do not exclude the possibility that a non-spherical solution, which we have not studied, has lower energy than the spherical hairy black hole.}  
The salient feature of the thermodynamic property of the hairy black hole solution is that not only the energy but also the chemical potential is a continuous function with respect to $Q$ (while the higher derivative can be discontinuous). This can be explicitly seen from our formula in the previous section where the energy difference is $O((q-q_c)^2)$ or higher. It is continuous but is not analytic because the higher derivative shows the discontinuity at $Q=Q_c$.
We further realize that it is a convex function (at least within the small charge limit we have studied). We therefore conclude that the convexity of the energy is restored by the existence of the hairy black hole solution. Since the convex function with $E(0)=0$ is superadditive (see Appendix~\ref{sec:conv_vs_spad}), the energy is superadditive with respect to charge.

Let us briefly discuss the nature of the phase transition from the AdS/CFT viewpoint. The big difference between the (extremal) Reissner-Nordstrom black hole and the hairy black hole is that the former does not have scalar hair. The scalar hair or the scalar condensation induced in the hairy black hole suggests that in the dual CFT, the scalar operator which is dual to $\phi$ (with conformal dimension $\Delta$) has a non-zero expectation value on the cylinder in this heavy state. It is distinct from the one-particle state that is dual to low-dimensional single trace operators with zero scalar expectation value. In the planer black hole (or black membrane), the analogous phase transition discussed here is referred to as a ``holographic superconductor" \cite{Hartnoll:2008vx}\cite{Hartnoll:2008kx} (or holographic superfluid), but we should point out that the phase transition here is about the properties of operators on the plane (or states in the cylinder).\footnote{Schematically, one may say that while the Reissner-Nordstrom black hole is dual to $\mathrm{Tr}(\Phi^m)$ with $m \sim O(N^2)$, the hairy black hole obtained as adding soliton should look like $\mathrm{Tr}(\Phi^m) (\mathrm{Tr}(\Phi^2))^l$. It may be interesting to study the implication of the phase transition in terms of the recent studies of black hole operators in CFTs \cite{Agarwal:2020zwm}\cite{Choi:2021rxi}. }

\section{Discussions}
\label{sec:discussion}

In this paper, we have studied the convexity of the lowest energy of the Einstein-Maxwell-charged scalar system in AdS under the presence of a hairy black hole. In the small charge limit, the existence of the hairy black hole solution restores the would-be violated convexity due to the phase transition between a charged soliton and a Reissner-Nordstrom black hole. It should be interesting to perform (numerical) analysis in the larger charge regime and examine if the convexity remains true.

Indeed, there has been some interest in studying the lowest conformal dimensions of conformal field theories as a function of the charge, in particular in the large charge expansion. Recent studies show that the large charge behavior of the generic conformal field theories should be $E(Q) \sim Q^{\frac{d}{d-2}}$ \cite{Hellerman:2015nra}\cite{Jafferis:2017zna}. We can easily check that the extremal Reissner-Nordstrom black hole satisfies this scaling. It was also numerically verified in charged soliton (with $\Delta=2$ in $d=3$) in \cite{delaFuente:2020yua}. It will be interesting to verify the behavior in the large charge hairy black holes with arbitrary $\Delta$. Note that in some supergravity, they found BPS hairy black hole solutions where $E(Q) \propto Q$ rather than $Q^{\frac{d}{d-2}}$ scaling. Apparently, the presence of the moduli allows the linear behavior. It is generally believed that it should not exhibit such a behavior without supersymmetry, but there is no proof.\footnote{By introducing $|\phi|^4$ interaction with a suitable coefficient, we expect that we can adjust the $Q^2$ term in $E(Q)$ in the charged soliton solutions so that it has a linear behavior in the small charge limit.}  In relation, it is highly doubtful but is not proved if there is any solution whose scaling behavior is in between $Q^\alpha$ where $ 1<\alpha < \frac{d}{d-2}$. It is an interesting study if we can or cannot engineer such solutions in gravity.

The study of the large charge behavior should shed light on the analytic property of large charge expansions. From the effective field theory viewpoint, it was not obvious if the large charge expansion is converging (down to $Q \sim 0$), has a finite radius of convergence, or is just an asymptotic expansion. The extreme Reissner-Nordstrom solution tells that the function is smooth for the entire range of charge, but the large charge expansion has a finite radius of convergence.\footnote{One can see that there is a branch cut in the negative $Q$ in $E_{RN}(Q)$, so one can read the radius of convergence in the positive $Q\sim \frac{1}{\kappa} $, which can be verified against d'Alembert's ratio test.} In this paper, we have shown that the hairy black hole solution shows a continuous phase transition, at which the function is not analytic. Whether the large charge expansion breaks down much above this phase transition point should be studied further in detail.

In this paper, we have studied charged solitons and hairy black holes under the Dirichlet boundary condition at the AdS boundary. It is an interesting question to allow more general boundary conditions such as the Robin boundary condition to study the charged solitons and hairy black holes \cite{Gentle:2011kv}\cite{Harada:2023cfl}. In the dual field theory perspective, it would correspond to double trace deformations \cite{Witten:2001ua}\cite{Berkooz:2002ug}. The stability condition may change and whether the convexity remains to hold is a non-trivial question to be addressed.

Our analysis of the charged solitons and hairy black holes was limited to $d \ge 3$ in this paper. For $d=2$, it is necessary to consider scalar hair around the charged BTZ black hole, and (extremal) hairy black holes can be drastically different from higher dimensions. The charge convexity conjecture in \cite{Aharony:2021mpc} is also restricted to $d > 2$, and it remains an open question if similar arguments can be established for $d=2$ CFT. It would be interesting if this situation could be approached from three-dimensional gravity duals.

Finally, it may be an interesting question to address the higher derivative gravitational corrections to the small charge (hairy) black holes and solitons. To satisfy the weak gravity conjecture in the Minkowski space-time, it is often suggested \cite{Cheung:2018cwt}\cite{Hamada:2018dde} that the higher derivative corrections make the extremal Reissner-Nordstrom black hole lighter. If the same mechanism is at work in the AdS, it may conflict with the convexity conjecture. The competition then may lead to new constraints on effective gravitational field theory in AdS.

\section*{Acknowledgments}

The authors would like to thank Yoshihiko Abe, Nick Dorey, and Toshifumi Noumi for fruitful discussions.
T.I.~is supported in part by JSPS KAKENHI Grant Number 19K03871. Y.N.~is in part supported by JSPS KAKENHI Grant Number 21K03581.

The authors thank the Yukawa Institute for Theoretical Physics at Kyoto University, where this work was initiated during the YITP workshop ``Strings and Fields 2023".

\appendix

\section{Convexity vs Superadditivity}
\label{sec:conv_vs_spad}
In this appendix, we collect some mathematical facts about convex functions in relation to superadditivity.

When a function $f(x)$ is convex between $x_1 \le x \le x_2$, for all $ 0 \le \lambda \le 1$, $f(x)$ satisfies the inequality
\begin{align}
\lambda f(x_1) + (1-\lambda) f(x_2) \ge f(\lambda x_1 + (1-\lambda) x_2) \ . \label{convexdef}
\end{align}

When $f(x)$ is twice differentiable, $f''(x) \ge 0$ in a segment implies the convexity in the same segment. The proof is based on Taylor's theorem 
\begin{align}
f(x) = f(x_0) + f'(x_0)(x-x_0) + \frac{f''(x_*)}{2} (x-x_0)^2 \ ,  
\end{align}
where $x_0 \le x_* \le x$ . By noting $f''(x_*) \ge 0$ and  setting $x_0 = \lambda x_1 + (1-\lambda) x_2$ with $x = x_1$ and $x= x_2$, we have
\begin{align}
f(x_1) &\ge f(x_0) + f'(x_0) (1-\lambda) (x_1-x_2) \ , \cr
f(x_2) & \ge f(x_0) + f'(x_0) \lambda (x_2-x_1)  \ .
\end{align}
Adding $\lambda$ times the first line and $(1-\lambda)$ times the second line, we obtain \eqref{convexdef}. 

When $f(x)$ is convex in $x \ge 0$ and $f(0)=0$, $f(x)$ is superadditive (i.e. $f(a) + f(b) \le f(a+b)$).  To show this, we first set $x_2=0$ in \eqref{convexdef} to obtain $f(\lambda x_1) \le \lambda f(x_1)$. Then, noting $ f(a) = f(\frac{a}{a+b} (a+b)) \le \frac{a}{a+b} f(a+b)$, we obtain
\begin{align}
f(a) + f(b) \le \frac{a}{a+b} f(a+b)  + \frac{b}{a+b} f(a+b) = f(a+b) \ . 
\end{align}

Let $g_1(x)$ and $g_2(x)$ be two twice differentiable convex functions with $g_1(0) = g_2(0) = 0$, $g_1(x) \le g_2(x)$ when $0 \le x \le x_*$, and $g_2(x) \le g_1(x)$ when $ x_* \le x$. Let $f(x) = \mathrm{min}[g_1(x), g_2(x)]$ for $x \ge 0$. Generically, $f(x)$ is not convex, but we will show that it is superadditive. 

When $a,b \ge x_*$ or $a+b \le x_*$, the convexity of $g_1(x)$ or $g_2(x)$ immediately implies the claim. The nontrivial case is when $a, b \le x_*$ but $a+b \ge x_*$, or when $a \ge  x_* , b \le x_*$. In the former case, 
\begin{align}
f(a+b) - f(a) - f(b) = g_2(a+b) - g_1(a) - g_1(b) \ge g_2(a+b) - g_2(a) - g_2(b) \ge 0 \ ,
\end{align}
and in the latter case,
\begin{align}
f(a+b) - f(a) - f(b) = g_2(a+b) - g_2(a) - g_1(b) \ge g_2(a+b) - g_2(a) - g_2(b) \ge 0 \ ,
\end{align}
so the claim holds in all cases.

\section{Small charge extremal Reissner-Nordstrom-AdS black hole}
\label{sec:small_charge_RN_derivation}

We want to express the energy and chemical potential \eqref{eq_m_mu_RN} of the extremal Reissner-Nordstrom black hole as functions of the charge. At $T=0$, from $f'(r_h)=0$, the horizon radius and charge are related as
\begin{align}
q = \frac{r_h^{d-2}}{(d-2)\kappa} \sqrt{(d-1)(d-2+d \, r_h^2)} \ .
\label{eq_extremal_rh}
\end{align}
When the charge $q$ is small, this equation can be inverted by a series expansion as
\begin{align}
r_h^2 = \widetilde{Q}^{\frac{2}{d-2}} - \frac{d}{(d-2)^2} \widetilde{Q}^{\frac{4}{d-2}} + \frac{d^2(d+1)}{2(d-2)^4} \widetilde{Q}^{\frac{6}{d-2}} - \frac{d^4(d+2)}{3(d-2)^6} \widetilde{Q}^{\frac{8}{d-2}} + \cdots \ ,
\end{align}
where $\widetilde{Q}$ is defined in \eqref{eq_widetildeQ}. Substituting the above expansion into $E$ and $\mu$ in \eqref{eq_m_mu_RN} and \eqref{eq_EQ_by_mq}, we obtain \eqref{eq_RN_Emu_Qseries}.

\section{Explicit solution for the charged soliton}
\label{sec:soliton_derivation}
In this appendix, we present the explicit form of the solution of the equations of motion describing a horizonless charged soliton in the small charge expansion. We begin with the spherically symmetric ansatz for the static solution of the Einstein-Maxwell-charged scalar system,
\begin{align}
ds^2 &= - f(r) dt^2 + g(r) dr^2 + r^2 d\Omega_{d-1}^2 \ , \cr
A_\mu dx^\mu & = A_t(r) dt \ , \cr
\phi & = \phi(r) \ ,
\end{align}
where $d\Omega_{d-1}^2$ is the line element of a unit $S^{d-1}$. We have chosen the gauge so that the complex scalar field $\phi$ is a real function of $r$.
Substituting the ansatz into the equations of motion of \eqref{action}, we obtain a set of equations to be solved:
\begin{align}
\frac{d-1}{2} r f' + \left( \frac{(d-1)(d-2)}{2} (1-g) + \Lambda r^2 g - \kappa^2 r^2 \phi'^2 + \kappa^2 m_\phi^2 r^2 g \phi^2 \right) f & \cr
+ \frac{\kappa^2}{2} r^2 A_t'^2 - \kappa^2 e^2 r^2 g A_t^2 \phi^2 &= 0 \ , \cr
\frac{d-1}{2} r g' - \left( \frac{(d-1)(d-2)}{2} (1-g) + \Lambda r^2 g + \kappa^2 r^2 \phi'^2 + \kappa^2 m_\phi^2 r^2 g \phi^2 \right) g & \cr
- \frac{\kappa^2 g}{2 f} r^2 A_t'^2 - \frac{\kappa^2 e^2 r^2 g^2}{f} A_t^2 \phi^2 &= 0 \ , \cr
A_t'' + \left( \frac{d-1}{r} - \frac{f'}{2 f} - \frac{g'}{2 g} \right) A_t' - 2 e^2 g \phi^2 A_t &= 0 \ , \cr
\phi'' + \left( \frac{d-1}{r} + \frac{f'}{2f} - \frac{g'}{2g} \right) \phi' + \left( \frac{e^2 A_t^2}{f} - m_\phi^2 \right) g \phi &= 0 \ .
\end{align}
Here, $'$ denotes the derivative with respect to $r$.

Around the background with $\phi=0$, we will solve these equations in perturbative series with respect to a small parameter $\epsilon$ as
\begin{align}
f(r) &= f_{(0)}(r) + f_{(2)}(r)\epsilon^2 + f_{(4)}(r)\epsilon^4 + \cdots \ , \cr
g(r) &= g_{(0)}(r) + g_{(2)}(r)\epsilon^2 + g_{(4)}(r)\epsilon^4 + \cdots \ , \cr
A_t(r) &= A_{t(0)}(r) + A_{t(2)}(r)\epsilon^2 + A_{t(4)}(r)\epsilon^4 + \cdots \ , \cr
\phi(r) &= \phi_{(1)}(r) \epsilon + \phi_{(3)}(r)\epsilon^3 + \phi_{(5)}(r)\epsilon^5 + \cdots \ .
\end{align}
The strategy is to first solve the scalar equation at $O(\epsilon)$, where the integration constants are fixed by the asymptotic behavior. Then, at $O(\epsilon^2)$, we first solve $g$ and $f$. Finally, we solve $A_t$. One integration constant appearing in $A_t$ is not fixed at this point but will be fixed by studying the regularity of the scalar function in $O(\epsilon^3)$. In this sense, the equations at $O(\epsilon^2)$ and $O(\epsilon^3)$ are bundled. Then, we continue to $O(\epsilon^4)$ for $f$, $g$, and $A_t$, and so on.

In the zeroth order in $\epsilon$, the background is the empty AdS spacetime with a constant gauge field specified by the chemical potential $\mu_{(0)}$,
\begin{align}
f_{(0)}(r) = 1 + r^2 \ , \qquad
g_{(0)}(r) = \frac{1}{1 + r^2} \ , \qquad
A_{t(0)}(r) = \mu_{(0)} \ ,
\end{align}
where we assume $\mu_{(0)}>0$. The empty AdS spacetime has zero charge.

At $O(\epsilon)$, we turn on the scalar field. The equation for $\phi_{(1)}$ reads
\begin{align}
\phi_{(1)}'' + \left( \frac{d-1}{r} + \frac{2 r}{1+r^2} \right) \phi_{(1)}' + \frac{1}{1+r^2} \left( \frac{e^2 \mu_{(0)}^2}{1+r^2} - \Delta(\Delta -d) \right) \phi_{(1)} &= 0 \ ,
\label{eq_phi_1st}
\end{align}
where we used $m_\phi^2 = \Delta(\Delta-d)$. At the center of the AdS $r=0$, this second-order differential equation has regular and diverging solutions. To solve this equation (without horizon), we impose regularity at $r=0$. In the AdS boundary $r \to \infty$, we impose the absence of the source discussed as follows.

In $r \to \infty$, the behavior of the scalar field (not specific to $O(\epsilon)$ but to any orders) is $\phi = c_0 ( r^{\Delta-d} + \cdots ) + c_1 ( r^{-\Delta} + \cdots )$, where $c_0$ and $c_1$ are constants. For the absence of the source in the boundary field theory, we require $c_0=0$. When $\Delta > d/2$, $r^{-\Delta}$ decays faster than $r^{\Delta-d}$. Hence, setting $c_0=0$ corresponds to keeping the subleading behavior of $\phi$. This case is called the standard quantization. When $\Delta < d/2$, $r^{-\Delta}$ decays slower than $r^{\Delta-d}$, but the leading behavior is normalizable for $\Delta$ in the range $d/2-1 < \Delta < d/2$ \cite{Balasubramanian:1998sn}. This case is called the alternative quantization. The border $\Delta=d/2$ is the case that the scalar mass saturates the Breitenlohner-Freedman (BF) bound \cite{Breitenlohner:1982bm}\cite{Breitenlohner:1982jf}, where we have $\phi = c_0 ( r^{-d/2} \log r + \cdots ) + c_1 ( r^{-d/2} + \cdots )$.

In the absence of the scalar field source $c_0=0$, the equation \eqref{eq_phi_1st} is solved at discrete values of $\mu_{(0)}$ satisfying
\begin{align}
\mu_{(0)} = \frac{\Delta + 2n}{e} \ ,
\end{align}
where $n$ is a non-negative integer. This is nothing but the normal mode frequency $\omega$ of the massive scalar field in the global AdS as $\omega = e \mu_{(0)} = \Delta + 2 n$. At this $\mu_{(0)}$, the regular solution to \eqref{eq_phi_1st} is given by the hypergeometric function as
\begin{align}
\phi_{(1)} = (1+r^2)^{\frac{\Delta}{2}+n} {}_2 F_1 \left( \frac{d}{2}+n , \Delta+n ; \frac{d}{2} ; -r^2 \right) \ ,
\label{eq_phi1_sol}
\end{align}
whose normalization is absorbed by $\epsilon$. Using this as the seed, we proceed to the higher orders in the perturbative series. In the following, we consider the solution with the lowest energy for a given $\Delta$, and hence we set $n=0$. For $n=0$, the above chemical potential and scalar field solution become
\begin{align}
\mu_{(0)} = \frac{\Delta}{e} \ , \qquad
\phi_{(1)} = \frac{1}{(1+r^2)^{\Delta/2}} \ .
\end{align}

For $O(\epsilon^2)$ and higher, we have solved these equations for integer values of $\Delta$ in various dimensions $d$. Some sample solutions are presented here.

For $d=4, \, \Delta=4 \ (m_\phi^2=0)$, we reproduce the results in \cite{Basu:2010uz}. We obtain
\begin{align}
f(r) &= 1+r^2 - \frac{8(3+3r^2+r^4)}{9(1+r^2)^3} \kappa^2 \epsilon^2 + \cdots \ , \cr
g(r) &= \frac{1}{1+r^2} + \frac{8r^2(3+r^2)}{9(1+r^2)^5} \kappa^2 \epsilon^2 + \cdots \ , \cr
A_t(r) &= \frac{4}{e} + \left( \frac{3 e}{14} - \frac{(3+3r^2+r^4)e}{6(1+r^2)^3} - \frac{32\kappa^2}{21e} \right) \epsilon^2 + \cdots \ , \cr
\phi(r) &= \frac{\epsilon}{(1+r^2)^2} + \cdots \ .
\end{align}
Here, we present the results up to $O(\epsilon^2)$, but the perturbative solutions can be obtained also in higher orders in $\epsilon$. Once the equations of motion are solved, the mass $m$, chemical potential $\mu$, and charge $q$ can be read off from the asymptotic behavior of the fields in $r \to \infty$ as
\begin{align}
f(r) &= r^2 + 1 + \cdots - \frac{m}{r^{d-2}} + \cdots \ , \cr
A_t(r) &= \mu + \cdots - \frac{q}{r^{d-2}} + \cdots \ .
\end{align}
These are chosen to agree with the notation of $m$, $\mu$, and $q$ used in the Reissner-Nordstrom black hole solution \eqref{eq_RN_solution}. For $d=4, \, \Delta=4$, we find
\begin{align}
m &= \frac{8 \kappa^2}{9} \epsilon^2 + \frac{(78336 \kappa^2 - 6767 e^2)\kappa^2}{39690} \epsilon^4 + \cdots \ , \cr
\mu &= \frac{4}{e} + \left( \frac{3e}{14} - \frac{32 \kappa^2}{21 e}\right) \epsilon^2 + \frac{122400480 \kappa^2 e^2 - 574944256 \kappa^4 - 6383817 e^4}{97796160 e} \epsilon^4 + \cdots \ , \cr
q &= \frac{e}{6} \epsilon^2 + \frac{(2658 \kappa^2 - 241 e^2)e}{6615} \epsilon^4 + \cdots \ ,
\end{align}
where we explicitly included the $O(\epsilon^4)$ contributions that are necessary for obtaining $m$ to $O(q^2)$. In the above expression, we have used the degrees of freedom to rescale $\epsilon$ to set the coefficient of $r^{-\Delta}$ in $\phi(r)$ in $r \to \infty$ to be exactly $\epsilon$ (no higher orders in $\epsilon$): $\phi(r) = \epsilon/r^4 + \cdots$. From this we can also read off the scalar condensate (vacuum expectation value), but we will not use it, so we omit it. Because we wish to express $m$ and $\mu$ perturbatively as functions of small $q$, we invert $q$ as
\begin{align}
\epsilon = \sqrt{\frac{6}{e}} \, q^{1/2} + \frac{2 \sqrt{6} (241 e^2 - 2658 \kappa^2)}{735 e^{3/2}} q^{3/2} + \cdots \ .
\end{align}
Substituting this into $m$ and $\mu$, we obtain
\begin{align}
m &= \frac{16 \kappa^2}{3 e} q + \frac{2 \kappa^2}{21} \left( 9 - \frac{64 \kappa^2}{e^2} \right) q^2 + \cdots \ , \cr
\mu &= \frac{4}{e} + \left( \frac{9}{7} - \frac{64 \kappa^2}{7 e^2} \right) q + \cdots \ .
\end{align}
Thus, we reproduce the results for the massless scalar in AdS$_5$ \cite{Basu:2010uz}.

We also provide some details for massive scalar in asymptotically AdS$_4$ $(d=3)$ with $\Delta=2$ $(m_\phi^2=-2)$ \cite{Gentle:2011kv}. We find
\begin{align}
f(r) &= 1+r^2 - \frac{r+(1+r^2) \tan^{-1} r}{r(1+r^2)} \kappa^2 \epsilon^2 + \cdots \ , \cr
g(r) &= \frac{1}{1+r^2} - \frac{r-(1+r^2) \tan^{-1} r}{r(1+r^2)^3} \kappa^2 \epsilon^2 + \cdots \ , \cr
A_t(r) &= \frac{2}{e} + \left( \frac{e(1+5r^2)}{8(1+r^2)} - \frac{\kappa^2}{2e} - \frac{e \tan^{-1} r}{2r} \right) \epsilon^2 + \cdots \ , \cr
\phi(r) &= \frac{\epsilon}{1+r^2} + \cdots \ .
\end{align}
From these, up to $O(\epsilon^4)$, we obtain\footnote{Note that our expansion parameter $\epsilon$ is different from \cite{Gentle:2011kv}, and therefore the coefficients in \eqref{m_mu_epsilonfn_d3Delta2} are different in higher orders in $\epsilon$. However, the ambiguities in the parametrization of $\epsilon$ cancel out when $m$ and $\mu$ are written as functions of $q$ as in \eqref{m_mu_qfn_d3Delta2}.}
\begin{align}
m &= \frac{\pi \kappa^2}{2} \epsilon^2 + \frac{\pi \kappa^2}{192} \left((65+4\pi^2)e^2 - 4 (61+4\pi^2)\kappa^2\right) \epsilon^4 + \cdots \ , \cr
\mu &= \frac{2}{e} + \left( \frac{5e}{8} - \frac{\kappa^2}{2 e}\right) \epsilon^2 + \frac{41 e^4(2\pi^2-15) + 4 \kappa^2 e^2 (661-93 \pi^2) + 4 \kappa^4 (98 \pi^2 - 821)}{384e} \epsilon^4 + \cdots \ , \cr
q &= \frac{\pi e}{4} \epsilon^2 + \frac{\pi e}{192} \left( e^2 (25+2\pi^2) - 4\kappa^2 (29+2\pi^2) \right) \epsilon^4 + \cdots \ .
\label{m_mu_epsilonfn_d3Delta2}
\end{align}
Rewriting $m$ and $\mu$ as functions of $q$, we find
\begin{align}
m &= \frac{2 \kappa^2}{e} q + \frac{\kappa^2}{4\pi} \left( 5 - \frac{4 \kappa^2}{e^2} \right) q^2 + \cdots \ , \cr
\mu &= \frac{2}{e} + \frac{1}{2\pi} \left( 5 - \frac{4 \kappa^2}{e^2} \right) q + \cdots \ .
\label{m_mu_qfn_d3Delta2}
\end{align}

We repeat the above calculations for various integer values of $\Delta$ in general dimensions $d$. Some results of $m$ and $\mu$ in small $q$ expansion are listed below.
\begin{itemize}
\item $d=3, \, \Delta=1$ (alternative quantization):
\begin{align}
m &= \frac{\kappa^2}{e} q + \frac{\kappa^2}{2\pi} \left( 1 + \frac{\kappa^2}{e^2} \right) q^2 + \cdots \ , \cr
\mu &= \frac{1}{e} + \frac{1}{\pi} \left( 1 + \frac{\kappa^2}{e^2} \right) q + \cdots \ .
\end{align}
\item $d=3, \, \Delta=2$:
\begin{align}
m &= \frac{2 \kappa^2}{e} q + \frac{\kappa^2}{4\pi} \left( 5 - \frac{4 \kappa^2}{e^2} \right) q^2 + \cdots \ , \cr
\mu &= \frac{2}{e} + \frac{1}{2\pi} \left( 5 - \frac{4 \kappa^2}{e^2} \right) q + \cdots \ .
\end{align}
\item $d=3, \, \Delta=3$:
\begin{align}
m &= \frac{3 \kappa^2}{e} q + \frac{7\kappa^2}{4\pi} \left( 1 - \frac{3 \kappa^2}{e^2} \right) q^2 + \cdots \ , \cr
\mu &= \frac{3}{e} + \frac{7}{2\pi} \left( 1 - \frac{3 \kappa^2}{e^2} \right) q + \cdots \ .
\end{align}
\item $d=4, \, \Delta=2$ (BF bound):
\begin{align}
m &= \frac{8 \kappa^2}{3 e} q + \frac{2 \kappa^2}{9} q^2 + \cdots \ , \cr
\mu &= \frac{2}{e} + \frac{1}{3} q + \cdots \ .
\end{align}
\item $d=4, \, \Delta=3$:
\begin{align}
m &= \frac{4 \kappa^2}{e} q + \frac{8 \kappa^2}{15} \left( 1 - \frac{3 \kappa^2}{e^2} \right) q^2 + \cdots \ , \cr
\mu &= \frac{3}{e} + \frac{4}{5} \left( 1 - \frac{3 \kappa^2}{e^2} \right) q + \cdots \ .
\end{align}
\item $d=4, \, \Delta=4$:
\begin{align}
m &= \frac{16 \kappa^2}{3 e} q + \frac{2 \kappa^2}{21} \left( 9 - \frac{64 \kappa^2}{e^2} \right) q^2 + \cdots \ , \cr
\mu &= \frac{4}{e} + \left( \frac{9}{7} - \frac{64 \kappa^2}{7 e^2} \right) q + \cdots \ .
\end{align}
\item $d=5, \, \Delta=2$ (alternative quantization):
\begin{align}
m &= \frac{3\kappa^2}{e} q + \frac{\kappa^2}{8\pi} \left( 1 + \frac{2 \kappa^2}{e^2} \right) q^2 + \cdots \ , \cr
\mu &= \frac{2}{e} + \frac{1}{6 \pi} \left( 1 + \frac{2 \kappa^2}{e^2} \right) q + \cdots \ .
\end{align}
\item $d=5, \, \Delta=3$:
\begin{align}
m &= \frac{9 \kappa^2}{2e} q + \frac{\kappa^2}{16\pi} \left( 14 - \frac{27 \kappa^2}{e^2} \right) q^2 + \cdots \ , \cr
\mu &= \frac{3}{e} + \frac{1}{12\pi} \left( 14 - \frac{27 \kappa^2}{e^2} \right) q + \cdots \ .
\end{align}
\item $d=5, \, \Delta=4$:
\begin{align}
m &= \frac{6 \kappa^2}{e} q + \frac{3\kappa^2}{16\pi} \left( 11 - \frac{72 \kappa^2}{e^2} \right) q^2 + \cdots \ , \cr
\mu &= \frac{4}{e} + \frac{1}{4\pi} \left( 11 - \frac{72 \kappa^2}{e^2} \right) q + \cdots \ .
\end{align}
\end{itemize}

By comparing such results for integer values of $\Delta$ in general dimensions $d$, we find that the general expression for $m$ and $\mu$ of the soliton up to $O(q^2)$ is given by
\begin{align}
m &= \frac{2(d-2)\Delta\kappa^2}{(d-1)e} q + \frac{(d-2)\kappa^2 \, \Gamma(\Delta)^2 \, \Gamma(2\Delta+1-d/2)}{(d-1)\Gamma(d/2) \, \Gamma(2\Delta) \, \Gamma(\Delta+1-d/2)^2} \left( 1 - \frac{(d-2)\Delta^2(\Delta-d/2)}{(d-1)(\Delta-d/4)} \frac{\kappa^2}{e^2} \right) q^2 \ , \cr
\mu &= \frac{\Delta}{e} + \frac{\Gamma(\Delta)^2 \, \Gamma(2\Delta+1-d/2)}{\Gamma(d/2) \, \Gamma(2\Delta) \, \Gamma(\Delta+1-d/2)^2}\left( 1 - \frac{(d-2)\Delta^2(\Delta-d/2)}{(d-1)(\Delta-d/4)} \frac{\kappa^2}{e^2} \right) q \ .
\end{align}

We note that coefficients  diverge when $\Delta=d/4$, which can be realized only in $d=3$ at $\Delta=3/4$ (in $d \ge 4$, we always have $\Delta>d/2-1\ge d/4$). This expansion formula, deduced from the results of integer $\Delta$, hence might not be suitable for $\Delta\le 3/4$ in $d=3$. At the end of section 2, we have discussed the same issue from the viewpoint of charge convexity. To avoid subtleties, in this paper we assume $\Delta>3/4$ when $d=3$.

\section{Construction of small charge hairy black holes}
\label{sec:hairy_derivation}
The critical coupling $e=e_c$ is where the extremal Reissner-Nordstrom black hole and soliton have the same energy for the same charge. Here, near $e=e_c$, we evaluate the energy of the hairy black hole as a non-interacting mix of the extremal Reissner-Nordstrom black hole and soliton. This method was successfully employed for massless scalar in $d=4$ in \cite{Dias:2011tj}, and here we generalize the calculations to other $\Delta$ and $d$.

\subsection{Construction in $e<e_c$}

In $e<e_c$, the Reissner-Nordstrom black hole has a lower energy than the soliton around $Q=Q_c$. At $Q=Q_c$, we start to mix small charge soliton contributions to the Reissner-Nordstrom black hole to evaluate the energy of a hairy black hole. We will see that the Reissner-Nordstrom black hole is the lowest energy state in $Q < Q_c$, while the hairy black hole is the lowest energy state in $Q>Q_c$.

In $e < e_c$, we express the small difference of the coupling $e$ from $e_c$ by introducing a small parameter $\theta$ as
\begin{align}
e^2 = e_c^2 (1 - \theta) \ .
\end{align}

From \eqref{eq_extremal_rh}, the chemical potential and charge of the extremal Reissner-Nordstrom black hole depends on small $r_h$,
\begin{align}
\mu_{RN} &=  \frac{1}{\kappa} \sqrt{\frac{d-1}{d-2}} \left( 1 + \frac{d}{2(d-2)}  r_h^2 + O(r_h^4) \right) \  , \cr
q_{RN} &=  \frac{r_h^{d-2}}{\kappa} \sqrt{\frac{d-1}{d-2}} \left( 1 + \frac{d}{2(d-2)}  r_h^2 + O(r_h^4) \right) \ .
\end{align}
To mix the Reissner-Nordstrom black hole and soliton, we parametrize $r_h$ and the charge of the soliton as perturbative series of $\theta$ so that they can be compared at each order in the perturbative series. In $e<e_c$, we introduce the ansatz that small soliton contributions are put on top of the Reissner-Nordstrom black hole. Hence, we devise the perturbative series so that $\mu_{RN}$ and $q_{RN}$ are larger than $\mu_{sol}$ and $q_{sol}$.

\subsubsection{$d=3$}

To mix the Reissner-Nordstrom black hole and soliton for $d=3$ in $e<e_c$, we take $q_{RN} \sim r_h \sim \theta^{1/2}$. We use the following ansatz for $r_h$,
\begin{align}
r_h = a_1 \theta^{1/2} + a_2 \theta + a_3 \theta^{3/2} + a_4 \theta^2 + \cdots \ ,
\end{align}
where we have explicitly included the terms necessary for calculating the hairy black hole energy up to $O(q^2)$. With this ansatz, $\mu_{RN}$, $q_{RN}$, and $m_{RN}$ of the Reissner-Nordstrom black hole can be expanded in powers of $\theta$ as
\begin{align}
\mu_{RN} &= \frac{\sqrt{2}}{\kappa} + \frac{3 a_1^2}{\sqrt{2} \kappa} \theta + \frac{3 \sqrt{2} a_1 a_2}{\kappa} \theta^{3/2} + \frac{3( 3 a_1^4 -4 a_2^2 - 8 a_1 a_3)}{4\sqrt{2} \kappa} \theta^{2} + \frac{- 9 a_1^3 a_2 + 6 a_2 a_3 + 6 a_1 a_4}{\sqrt{2} \kappa} \theta^{5/2} + \cdots \ , \cr
q_{RN} &= \frac{\sqrt{2} a_1}{\kappa} \theta^{1/2} + \frac{\sqrt{2} a_2}{\kappa} \theta + \frac{\sqrt{2} (3 a_1^3 + 2 a_3)}{2 \kappa} \theta^{3/2} + \frac{\sqrt{2} (9 a_1^2 a_2  + 2 a_4)}{2 \kappa} \theta^2 + \cdots \ , \cr
m_{RN} &= 2 a_1 \theta^{1/2} + 2 a_2 \theta + 2(2 a_1^3 + a_3) \theta^{3/2} + (6 a_1^2 a_2 + a_4) \theta^2 + \cdots \ .
\end{align}
To balance the chemical potential, we take the ansatz for the charge of the soliton as
\begin{align}
q_{sol} = b_1 \theta + b_2 \theta^{3/2} + b_3 \theta^2 + \cdots \ .
\end{align}
Here, the leading behavior of the soliton is chosen as $q_{sol} \sim \theta$ in order to balance the chemical potential. Then, the chemical potential and mass of the soliton can be expressed as
\begin{align}
\mu_{sol} &= \frac{\sqrt{2}}{\kappa} + \left( \frac{1}{\sqrt{2}\kappa} + \frac{3 b_1 \, \Gamma(\Delta) \, \Gamma(2 \Delta - 3/2)}{2^{2\Delta-1} \, \Gamma(\Delta+1/2) \, \Gamma(\Delta-1/2)^2} \right) \theta + \frac{3 b_2 \, \Gamma(\Delta) \, \Gamma(2 \Delta - 3/2)}{2^{2\Delta-1} \, \Gamma(\Delta+1/2) \, \Gamma(\Delta-1/2)^2} \theta^{3/2} \cr 
& \qquad + \left( \frac{3}{4 \sqrt{2} \kappa} + \frac{(3 b_3 + 2 b_1 (3-2\Delta)) \Gamma(\Delta) \, \Gamma(2 \Delta - 3/2)}{2^{2\Delta-1} \, \Gamma(\Delta+1/2) \, \Gamma(\Delta-1/2)^2} \right) \theta^2 + \cdots \ , \cr
m_{sol} &= \sqrt{2} b_1 \kappa \theta + \sqrt{2} b_2 \kappa \theta^{3/2} + \left( \frac{(b_1 + 2 b_3) \kappa}{\sqrt{2}} + \frac{3 b_1^2 \kappa^2 \, \Gamma(\Delta) \, \Gamma(2 \Delta - 3/2)}{2^{2 \Delta} \, \Gamma(\Delta+1/2) \, \Gamma(\Delta-1/2)^2} \right) \theta^2 + \cdots \ ,
\end{align}
where $b_1>0$ is assumed so that $m_{sol}>0$. Note that, with the above ansatz, the charge of the hairy black hole has the series expansion of the form
\begin{align}
q = q_{RN} + q_{sol} = \frac{\sqrt{2} a_1}{\kappa} \theta^{1/2} + \left( \frac{\sqrt{2} a_2}{\kappa} + b_1 \right) \theta + \cdots \ .
\label{eq_qmix_d3_e<ec}
\end{align}

At this point, we fix the redefinition ambiguities in the $\theta$ expansion by demanding $q \sim \theta^{1/2}$ with no higher $\theta$ corrections.\footnote{In \cite{Dias:2011tj}, a different choice to fix the ambiguities was made (i.e. $q_{sol}$ has no higher order $\theta$ corrections), but eventually, if we express $m_{hairy}$ as a function of $q$, the ambiguities will be gone.} Then, each term in the above expansion should satisfy
\begin{align}
q = \frac{\sqrt{2} a_1}{\kappa} \theta^{1/2} \ , \qquad
\frac{2 a_2}{\kappa} + b_1 = 0 \ ,
\end{align}
and so on for the higher orders involving $a_{3,4}$ and $b_{2,3}$. 
The balancing of the chemical potential $\mu_{RN} = \mu_{sol}$ also gives the equations
\begin{align}
\frac{3 a_1^2}{\sqrt{2} \kappa} &= \frac{1}{\sqrt{2}\kappa} + \frac{3 b_1 \, \Gamma(\Delta) \, \Gamma(2 \Delta - 3/2)}{2^{2\Delta-1} \, \Gamma(\Delta+1/2) \, \Gamma(\Delta-1/2)^2} \ , \cr
\frac{3 \sqrt{2} a_1 a_2}{\kappa} &= 
\frac{3 b_2 \, \Gamma(\Delta) \, \Gamma(2 \Delta - 3/2)}{2^{2\Delta-1} \, \Gamma(\Delta+1/2) \, \Gamma(\Delta-1/2)^2} \ ,
\end{align}
and so on for $a_{3}$ and $b_{3}$. From these equations, we can determine the coefficients $a_{1,2,3,4}$ and $b_{1,2,3}$. The upshot is that, in $q>q_c$, the mass of the hairy black hole $m_{hairy} = m_{RN} + m_{sol}$ is given by
\begin{align}
m_{hairy} = m_{RN} - \frac{3 \times 2^{2 \Delta - 5} \, \Gamma(\Delta-1/2)^2 \, \Gamma(\Delta+1/2)}{\Gamma(\Delta) \, \Gamma(2\Delta-3/2)} \kappa^4 (q^2 - q_c^2)^2 \ ,
\label{eq_m_hairy_ads4_app}
\end{align}
where
\begin{align}
q_c &= \frac{\sqrt{6}}{3 \kappa} \theta^{1/2} \ , \cr
m_{RN} &= \sqrt{2} \kappa q + \frac{\kappa^3 q^3}{2\sqrt{2}} + O(q^5) \ .
\end{align}
Recall we assume $\Delta > 3/4$. Then, from \eqref{eq_m_hairy_ads4_app}, we see that $m_{hairy} < m_{RN}$.
Note that we also find $b_1 \to 0$ in $q \to q_c$ from above.

In fact, we find $q_{sol}<0$ in $q<q_c$, and therefore the above expression cannot be applied in $q<q_c$, where $m_{sol}$ would be $m_{sol}<0$ if $m_{sol} = q_{sol} + \cdots$ is assumed. This means that we need to parametrize the soliton mass as $m_{sol} = |q_{sol}| + \cdots $. Then, repeating the calculations, we find that the mass of the hairy black hole in $q<q_c$ is given by
\begin{align}
m_{hairy} = m_{RN} + \frac{2^{2 \Delta - 1} \, \Gamma(\Delta-1/2)^2 \, \Gamma(\Delta+1/2)}{\Gamma(\Delta) \, \Gamma(2\Delta-3/2)} \kappa^2 (q_c^2 - q^2) \ .
\end{align}
Thus, $m_{hairy} >  m_{RN}$ in $q<q_c$ if $\Delta>3/4$. This is reasonable because in $q<q_c$ this construction attempts to approximate the hairy black hole as a non-interacting mix of the Reissner-Nordstrom black hole and soliton with the opposite charges.

\subsubsection{$d=4$}

In $d=4$, we use $q_{RN} \sim r_h^2 \sim \theta$. The ansatz for $r_h$ is
\begin{align}
r_h^2 = a_1 \theta + a_2 \theta^2 + \cdots \ .
\end{align}
With this ansatz, $\mu_{RN}$, $q_{RN}$, and $m_{RN}$ takes the form
\begin{align}
\mu_{RN} &= \frac{\sqrt{6}}{2 \kappa}+ \frac{\sqrt{6} a_1}{2\kappa} \theta + \frac{\sqrt{6}(2 a_2 - a_1^2)}{4 \kappa} \theta^2 + \cdots \ , \cr
q_{RN} &= \frac{\sqrt{6} a_1}{2 \kappa} \theta +  \frac{\sqrt{6}(a_1^2 + a_2)}{2 \kappa} \theta^2 + \cdots \ , \cr
m_{RN} &= 2 a_1 \theta + (3 a_1^2 + 2 a_2) \theta^2 + \cdots \ .
\end{align}
To match this, we use the ansatz for the charge of the soliton as
\begin{align}
q_{sol} = b_1 \theta + b_2 \theta^2 + \cdots \ .
\end{align}
The chemical potential and mass of the soliton are
\begin{align}
\mu_{sol} &= \frac{\sqrt{6}}{2\kappa} + \left( \frac{b_1 (\Delta-1)}{2 \Delta - 1} + \frac{\sqrt{6}}{4\kappa} \right) \theta +\left( \frac{(\Delta-1)(b_2 - b_1 (\Delta - 2))}{2 \Delta - 1} + \frac{3 \sqrt{6}}{16 \kappa} \right) \theta^2 + \cdots \ , \cr
m_{sol} &= \frac{2 \sqrt{6} \kappa b_1}{3} \theta + \frac{1}{3} \left( \sqrt{6} \kappa (b_1 + 2 b_2) + \frac{2 \kappa^2 b_1^2 (\Delta-1)}{2 \Delta - 1} \right) \theta^2 + \cdots \ ,
\end{align}
where $b_1>0$. The charge of the hairy black hole has the series expansion of the form
\begin{align}
q = q_{RN} + q_{sol} = \left( \frac{\sqrt{6} a_1}{2 \kappa} + b_1 \right) \theta + \left( \frac{\sqrt{6} (a_1^2+a_2)}{2 \kappa} + b_2 \right) \theta^2 + \cdots \ .
\label{eq_qmix_d4_e<ec}
\end{align}
Solving $\mu_{RN} = \mu_{sol}$ and fixing the redefinition ambiguities by demanding $q \sim \theta$ with no higher $\theta$ corrections, we can determine $a_{1,2}$ and $b_{1,2}$. The mass $m_{hairy} = m_{RN} + m_{sol}$ is obtained in $q>q_c$ as
\begin{align}
m_{hairy} = m_{RN} - \frac{2 (2 \Delta - 1)}{3 (3\Delta - 2)} \kappa^2 (q - q_c)^2 \ ,
\end{align}
where
\begin{align}
q_c &= \frac{\sqrt{6}}{4 \kappa} \theta \ , \cr
m_{RN} &= \frac{2\sqrt{6}}{3} \kappa q + \frac{2}{3} \kappa^2 q^2 \ .
\end{align}
In $q<q_c$, where $q_{sol}<0$, we find
\begin{align}
m_{hairy} = m_{RN} + \frac{4 \sqrt{6} (2 \Delta - 1) \kappa}{3 (3\Delta - 2)} (q_c - q) > m_{RN} \ .
\end{align}

\subsubsection{$d=5$}

In $d=5$, we use $q_{RN} \sim r_h^3 \sim \theta^{3/2}$. The ansatz for $r_h$ we choose is 
\begin{align}
r_h = a_1 \theta^{1/2} + a_2 \theta + a_3 \theta^{3/2} + a_4 \theta^2 + \cdots \ .
\end{align}
With this ansatz, $\mu_{RN}$, $q_{RN}$, and $m_{RN}$ can be parametrized as
\begin{align}
\mu_{RN} &= \frac{2}{\sqrt{3} \kappa} + \frac{5 a_1^2}{3 \sqrt{3} \kappa} \theta + \frac{10 a_1 a_2}{3 \sqrt{3} \kappa} \theta^{3/2} + \frac{5 (-5 a_1^4 + 12 a_2^2 + 24 a_1 a_3)}{36 \sqrt{3} \kappa} \theta^2 \cr
& \qquad +  \frac{5 (-5 a_1^3 a_2 + 6 a_2 a_3 + 6 a_1 a_4)}{9 \sqrt{3} \kappa} \theta^{5/2} + \cdots \ , \cr
q_{RN} &= \frac{2 a_1^3}{\sqrt{3} \kappa} \theta^{3/2} + \frac{2 \sqrt{3} a_1^2 a_2}{\kappa} \theta^2 + \frac{5 a_1^5 + 18 a_1 a_2^2 + 18 a_1^2 a_3}{3 \sqrt{3} \kappa} \theta^{5/2} \cr
& \qquad + \frac{25 a_1^4 a_2 + 6 a_2^3 + 36 a_1 a_2 a_3 + 18 a_1^2 a_4}{3 \sqrt{3} \kappa} \theta^3 + \cdots \ , \cr
m_{RN} &= 2 a_1^3 \theta^{3/2} + 6 a_1^2 a_2 \theta^2 + \left( \frac{8 a_1^5}{3} + 6 a_1 a_2^2 + 6 a_1^2 a_3 \right) \theta^{5/2} \cr
& \qquad + \left( \frac{40 a_1^4 a_2}{3} + 2 a_2^3 + 12 a_1 a_2 a_3 + 6 a_1^4 a_4 \right) \theta^3 + \cdots \ .
\end{align}
For the matching of the scaling of the above charge, we take the ansatz for the charge of the soliton as
\begin{align}
q_{sol} = b_1 \theta^{3/2} + b_2 \theta^2 + b_3 \theta^{5/2} + \cdots \ .
\end{align}
The chemical potential and mass of the soliton are
\begin{align}
\mu_{sol} &= \frac{2}{\sqrt{3} \kappa} + \frac{1}{\sqrt{3}\kappa} \theta + \frac{5 \times 2^{2-2\Delta} b_1 \, \Gamma(\Delta) \, \Gamma(2 \Delta - 5/2)}{3 \, \Gamma(\Delta+1/2) \, \Gamma(\Delta-3/2)^2} \theta^{3/2} \cr
& \qquad + \left(  \frac{\sqrt{3}}{4 \kappa} + \frac{5 \times 2^{2-2\Delta} b_2 \, \Gamma(\Delta) \, \Gamma(2 \Delta - 5/2)}{3 \, \Gamma(\Delta+1/2) \, \Gamma(\Delta-3/2)^2} \right) \theta^2 \cr
& \qquad +  \frac{2^{2-2\Delta} (5 b_3 - 2 b_1 (2 \Delta - 5)) \Gamma(\Delta) \, \Gamma(2 \Delta - 5/2)}{3 \, \Gamma(\Delta+1/2) \, \Gamma(\Delta-3/2)^2} \theta^{5/2} + \cdots \ , \cr
m_{sol} &= \sqrt{3} \kappa b_1 \theta^{3/2} + \sqrt{3} \kappa b_2 \theta^2 + \frac{\sqrt{3}}{2} \kappa (2 b_3 + b_1) \theta^{5/2} + \cdots \ ,
\end{align}
where $b_1>0$. The charge of the hairy black hole has the series expansion of the form
\begin{align}
q = q_{RN} + q_{sol} = \left( \frac{2 a_1^3}{\sqrt{3} \kappa} + b_1 \right) \theta^{3/2} + \left( \frac{2 \sqrt{3} a_1^2 a_2}{\kappa} + b_2 \right) \theta^2 + \cdots \ .
\label{eq_qmix_d5_e<ec}
\end{align}
We solve $\mu_{RN} = \mu_{sol}$ and $q \sim \theta^{3/2} $ for $a_{1,2,3,4}$ and $b_{1,2,3}$. The mass $m_{hairy} = m_{RN} + m_{sol}$ is then given in $q>q_c$ by
\begin{align}
m_{hairy} &= \sqrt{3} \kappa q + \left( \frac{\sqrt{3}}{2} \kappa q \theta - \frac{6 \sqrt{15}}{125} \theta^{5/2} \right) \cr
&= m_{RN} - \left( \frac{3^{5/6} \kappa^{5/3} q^{5/3}}{2^{5/3}} - \frac{\sqrt{3}}{2} \kappa q \theta + \frac{6 \sqrt{15}}{125} \theta^{5/2} \right) \ ,
\end{align}
where
\begin{align}
q_c &= \frac{6}{5 \sqrt{5} \kappa} \theta^{3/2} \ , \cr
m_{RN} &= \sqrt{3} \kappa q + \frac{3^{5/6} \kappa^{5/3} q^{5/3}}{2^{5/3}} \ .
\end{align}
In $q<q_c$, where $q_{sol}<0$, we find
\begin{align}
m_{hairy} = m_{RN} + 2 \sqrt{3} \kappa (q_c - q) > m_{RN} \ .
\end{align}

\subsubsection{$d \ge 6$}
In $e < e_c$, hairy black holes for $d \ge 6$ can be constructed in a similar way as $d=5$ and are of no qualitative difference. We first notice $q_{RN} \sim r_h^{d-2} \sim \theta^{(d-2)/2}$ (see \eqref{Qc_e_less_ec} and \eqref{eq_extremal_rh}). Then, we parametrize $r_h$ and $q_{sol}$ by series expansion in $\theta$ so that $q_{RN}$ and $q_{sol}$ have the same powers of $\theta$. To calculate the energy up to $O(q^2)$, for even $d$, we use the ansatz
\begin{align}
r_h^2 &= a_1 \theta + a_2 \theta^2 + \cdots + a_{d/2} \theta^{{d/2}} + \cdots \ , \cr
q_{sol} &= b_1 \theta^{(d-2)/2} + b_2 \theta^{(d-2)/2+1} + \cdots + b_{d/2} \theta^{d-2} + \cdots \ .
\end{align}
and for odd $d$, we alternatively use the ansatz
\begin{align}
r_h &= a_1 \theta^{1/2} + a_2 \theta + \cdots + a_{d-1} \theta^{{(d-1)/2}} + \cdots \ , \cr
q_{sol} &= b_1 \theta^{(d-2)/2} + b_2 \theta^{(d-2)/2+1/2} + \cdots + b_{d-2} \theta^{d-2} + \cdots \ .
\end{align}
For example, for $d=6$, we have
\begin{align}
r_h^2 &= a_1 \theta + a_2 \theta^2 + a_3 \theta^3 + \cdots \ , \cr
q_{sol} &= b_1 \theta^2 + b_2 \theta^3 + b_3 \theta^4 + \cdots \ .
\end{align}

As we have done above, we can determine parameters $a_i$ and $b_i$ 
by balancing the chemical potential $\mu_{RN} = \mu_{sol}$ and demanding the charge to be $q \sim \theta^{(d-2)/2}$ (with no higher order corrections). We can systematically calculate the energy of the hairy black hole for any $d \ge 6$. Here, we list some results.
\begin{itemize}
\item $d=6$: We have $q \sim \theta^2$. The phase transition at $q=q_c$ is continuous as
\begin{align}
m_{hairy} &= m_{RN} - \left( \frac{2^{3/2} \kappa^{3/2} q^{3/2}}{5^{3/4}} - \frac{2}{\sqrt{5}} \kappa q \theta + \frac{4}{27} \theta^3 \right) \cr
&= m_{RN} - \frac{9 \kappa^2}{20 \, \theta} (q - q_c)^2 + O \left( (q - q_c)^3 \right) \ ,
\end{align}
where the second line is the Taylor expansion around $q=q_c$, where $|q-q_c| \ll \theta^2$, and
\begin{align}
m_{RN} &= \frac{4}{\sqrt{5}} \kappa q + \frac{2^{3/2} \kappa^{3/2} q^{3/2}}{5^{3/4}} + \cdots \ , \cr
q_c &= \frac{2 \sqrt{5}}{9 \kappa} \theta^2 \ .
\end{align}
By considering the $O(q^2)$ terms as well, we can check the convexity (because $\Delta>2$) as
\begin{align}
\frac{d^2 m_{hairy}}{d q^2} = \frac{6(\Delta-1)(\Delta-2)^2 \kappa^2}{5(2 \Delta - 1)(2 \Delta - 3)} > 0 \ .
\end{align}
\item $d=7$: We have $q \sim \theta^{5/2}$. The phase transition at $q=q_c$ is continuous as
\begin{align}
m_{hairy} &= m_{RN} - \left( \frac{5^{7/10} \kappa^{7/5} q^{7/5}}{6^{7/10}} - \sqrt{\frac{5}{6}} \, \kappa q \theta + \frac{2 \times 5^{5/2}}{7^{7/2}} \theta^{7/2} \right) \cr
&= m_{RN} - \frac{7^{5/2} \kappa^2}{6 \times 5^{5/2} \theta^{3/2}} (q - q_c)^2 + O \left( (q - q_c)^3 \right) \ ,
\end{align}
where the second line is the Taylor expansion around $q=q_c$, where $|q-q_c| \ll \theta^{5/2}$, and
\begin{align}
m_{RN} &= \sqrt{\frac{10}{3}} \, \kappa q + \frac{5^{7/10} \kappa^{7/5} q^{7/5}}{6^{7/10}} + \cdots \ , \cr
q_c &= \frac{25 \sqrt{6}}{7^{5/2} \kappa} \theta^{5/2} \ .
\end{align}
By considering the $O(q^2)$ terms as well, we can check the convexity (because $\Delta>5/2$) as
\begin{align}
\frac{d^2 m_{hairy}}{d q^2} = \frac{7 \kappa^2 \, \Gamma(\Delta) \, \Gamma(2 \Delta-7/2)}{9 \times 2^{2\Delta-3} \, \Gamma(\Delta+1/2) \, \Gamma(\Delta-5/2)^2} > 0 \ .
\end{align}
\item $d=8$: We have $q \sim \theta^3$. The phase transition at $q=q_c$ is continuous as
\begin{align}
m_{hairy} &= m_{RN} - \left( \frac{6^{2/3} \kappa^{4/3} q^{4/3}}{7^{2/3}} - \sqrt{\frac{6}{7}} \, \kappa q \theta + \frac{27}{256} \theta^4 \right) \cr
&= m_{RN} - \frac{64 \kappa^2}{189 \, \theta^2} (q - q_c)^2 + O \left( (q - q_c)^3 \right) \ ,
\end{align}
where the second line is the Taylor expansion around $q=q_c$, where $|q-q_c| \ll \theta^3$, and
\begin{align}
m_{RN} &= 2 \sqrt{\frac{6}{7}} \, \kappa q + \frac{6^{2/3} \kappa^{4/3} q^{4/3}}{7^{2/3}} + \cdots \ , \cr
q_c &= \frac{9 \sqrt{42}}{128 \kappa} \theta^3 \ .
\end{align}
By considering the $O(q^2)$ terms as well, we can check the convexity (because $\Delta>3$) as
\begin{align}
\frac{d^2 m_{hairy}}{d q^2} = \frac{2(\Delta-1)(\Delta-2)(\Delta-3)^2 \kappa^2}{7(2 \Delta - 1)(2 \Delta - 3)} > 0 \ .
\end{align}
\end{itemize}

\subsection{Construction in $e>e_c$}

In $e>e_c$, the soliton has a lower energy than the Reissner-Nordstrom black hole around $Q=Q_c$. At $Q=Q_c$, we start to mix small charge Reissner-Nordstrom black hole contributions to the soliton to obtain a hairy black hole. We will see that the soliton is the lowest energy state in $Q < Q_c$, while the hairy black hole is the lowest energy state in $Q>Q_c$.

In $e > e_c$, we write the small difference of the coupling $e$ from $e_c$ as
\begin{align}
e^2 = e_c^2 (1 + \theta) \ .
\end{align}

\subsubsection{$d=3$}

In $d=3$ and $e>e_c$, we take $q_{RN} \sim r_h \sim \theta$ so that $q_{RN}$ is not more dominant than $q_{sol}$. The ansatz for $r_h$ can be taken as
\begin{align}
r_h = a_1 \theta + a_2 \theta^2 + \cdots \ .
\end{align}
With this ansatz, $\mu_{RN}$, $q_{RN}$, and $m_{RN}$ can be parametrized as
\begin{align}
\mu_{RN} &= \frac{\sqrt{2}}{\kappa} + \frac{3 a_1^2}{\sqrt{2} \kappa} \theta^2 + \cdots \ , \cr
q_{RN} &= \frac{\sqrt{2} a_1}{\kappa} \theta + \frac{\sqrt{2} a_2}{\kappa} \theta^2 + \cdots \ , \cr
m_{RN} &= 2 a_1 \theta + 2 a_2 \theta^2 + \cdots \ .
\end{align}
To match this, we use the ansatz for the charge of the soliton as
\begin{align}
q_{sol} = b_1 \theta + b_2 \theta^2 + \cdots \ .
\end{align}
The chemical potential and mass of the soliton are then
\begin{align}
\mu_{sol} &= \frac{\sqrt{2}}{\kappa} + \left( - \frac{1}{\sqrt{2}\kappa} + \frac{3 b_1 \, \Gamma(\Delta) \, \Gamma(2 \Delta - 3/2)}{2^{2\Delta-1} \, \Gamma(\Delta+1/2) \, \Gamma(\Delta-1/2)^2} \right) \theta + \cr 
& \qquad + \left( \frac{3}{4 \sqrt{2} \kappa} + \frac{(3 b_2 + 2 b_1 (2\Delta-3)) \Gamma(\Delta) \, \Gamma(2 \Delta - 3/2)}{2^{2\Delta-1} \, \Gamma(\Delta+1/2) \, \Gamma(\Delta-1/2)^2} \right) \theta^2 + \cdots \ , \cr
m_{sol} &= \sqrt{2} b_1 \kappa \theta + \left( \frac{(2 b_2 - b_1) \kappa}{\sqrt{2}} + \frac{3 b_1^2 \kappa^2 \, \Gamma(\Delta) \, \Gamma(2 \Delta - 3/2)}{2^{2 \Delta} \, \Gamma(\Delta+1/2) \, \Gamma(\Delta-1/2)^2} \right) \theta^2 + \cdots \ ,
\end{align}
where $b_1>0$. The charge of the hairy black hole takes the form
\begin{align}
q = q_{RN} + q_{sol} = \left( \frac{\sqrt{2} a_1}{\kappa} + b_1 \right) \theta + \left( \frac{\sqrt{2} a_2}{\kappa} + b_2 \right) \theta^2 + \cdots \ .
\label{eq_qmix_d3_e>ec}
\end{align}
We solve $\mu_{RN} = \mu_{sol}$ and $q \sim \theta$ for $a_{1,2}$ and $b_{1,2}$. The mass $m_{hairy} = m_{RN} + m_{sol}$ is then given in $q>q_c$ as
\begin{align}
m_{hairy} &= \sqrt{2} \kappa q - \frac{ 2^{2\Delta-3} \, \Gamma(\Delta-1/2)^2 \, \Gamma(\Delta+1/2)}{3 \, \Gamma(\Delta) \, \Gamma(2\Delta-3/2)} \theta^2 \cr
&= m_{sol} - \frac{3 \, \Gamma(\Delta) \, \Gamma(2\Delta-3/2)}{2^{2 \Delta} \, \Gamma(\Delta-1/2)^2 \, \Gamma(\Delta+1/2)} \kappa^2 (q - q_c)^2 \ ,
\end{align}
where
\begin{align}
q_c &= \frac{2^{2 \Delta - 3/2} \, \Gamma(\Delta-1/2)^2 \, \Gamma(\Delta+1/2)}{3 \kappa \, \Gamma(\Delta) \, \Gamma(2\Delta-3/2)} \theta \ , \cr
m_{sol} &= \sqrt{2} \kappa q + \left(\frac{3 \, \Gamma(\Delta) \, \Gamma(2\Delta-3/2)}{2^{2 \Delta} \, \Gamma(\Delta-1/2)^2 \, \Gamma(\Delta+1/2)} \kappa^2 q^2 - \frac{\kappa q \theta}{\sqrt{2}} \right) \ .
\end{align}
In $q<q_c$, we find $q_{RN}<0$, and hence we repeat the calculations by using the expression of the mass to be $m_{RN} = 2 |a_1| \theta + \cdots$. Then, we obtain
\begin{align}
m_{hairy} = m_{sol} + 2 \sqrt{2} \kappa (q_c - q) > m_{sol} \ .
\end{align}

\subsubsection{$d=4$}

In $d=4$, we use $q_{RN} \sim r_h^2 \sim \theta$. The ansatz for $r_h$ is
\begin{align}
r_h^2 = a_1 \theta + a_2 \theta^2 + \cdots \ .
\end{align}
With this ansatz, $\mu_{RN}$, $q_{RN}$, and $m_{RN}$ are given by
\begin{align}
\mu_{RN} &= \frac{\sqrt{6}}{2 \kappa}+ \frac{\sqrt{6} a_1}{2\kappa} \theta + \frac{\sqrt{6}(2 a_2 - a_1^2)}{4 \kappa} \theta^2 + \cdots \ , \cr
q_{RN} &= \frac{\sqrt{6} a_1}{2 \kappa} \theta +  \frac{\sqrt{6}(a_1^2 + a_2)}{2 \kappa} \theta^2 + \cdots \ , \cr
m_{RN} &= 2 a_1 \theta + (3 a_1^2 + 2 a_2) \theta^2 + \cdots \ .
\end{align}
To match this, we use the ansatz for the charge of the soliton as
\begin{align}
q_{sol} = b_1 \theta + b_2 \theta^2 + \cdots \ .
\end{align}
The chemical potential and mass of the soliton are
\begin{align}
\mu_{sol} &= \frac{\sqrt{6}}{2\kappa} + \left( \frac{b_1 (\Delta-1)}{2 \Delta - 1} - \frac{\sqrt{6}}{4\kappa} \right) \theta +\left( \frac{(\Delta-1)(b_1 (\Delta - 2) + b_2)}{2 \Delta - 1} + \frac{3 \sqrt{6}}{16 \kappa} \right) \theta^2 + \cdots \ , \cr
m_{sol} &= \frac{2 \sqrt{6} \kappa b_1}{3} \theta + \frac{1}{3} \left( \sqrt{6} \kappa (2 b_2 - b_1) + \frac{2 \kappa^2 b_1^2 (\Delta-1)}{2 \Delta - 1} \right) \theta^2 + \cdots \ .
\end{align}
The charge of the hairy black hole is
\begin{align}
q = q_{RN} + q_{sol} = \left( \frac{\sqrt{6} a_1}{2 \kappa} + b_1 \right) \theta + \left( \frac{\sqrt{6} (a_1^2+a_2)}{2 \kappa} + b_2 \right) \theta^2 + \cdots \ .
\label{eq_qmix_d4_e>ec}
\end{align}
We solve $\mu_{RN} = \mu_{sol}$ and $q \sim \theta$ for $a_{1,2}$ and $b_{1,2}$. Then, we obtain the mass $m_{hairy} = m_{RN} + m_{sol}$ as a function of $q$. The result is, in $q>q_c$,
\begin{align}
m_{hairy} = m_{sol} - \frac{2 (\Delta - 1)^2}{3 (2\Delta - 1)(3\Delta - 2)} \kappa^2 (q - q_c)^2 \ ,
\end{align}
where
\begin{align}
q_c &= \frac{\sqrt{6} (2 \Delta - 1)}{4 (\Delta - 1) \kappa} \theta \ , \cr
m_{sol} &= \frac{2\sqrt{6}}{3} \kappa q + \left(\frac{2(\Delta-1)}{3 (2 \Delta - 1)} \kappa^2 q^2 - \frac{\sqrt{6}}{3} q \theta \right) \ .
\end{align}
In $q<q_c$, where $q_{RN}<0$, we find
\begin{align}
m_{hairy} = m_{sol} + \frac{4 \sqrt{6} (\Delta - 1) \kappa}{3 (3\Delta - 2)} (q_c - q) > m_{sol} \ .
\end{align}

\subsubsection{$d=5$}

In $d=5$, we use $q_{RN} \sim r_h^3 \sim \theta^{3/2}$. The ansatz for $r_h$ we choose is 
\begin{align}
r_h = a_1 \theta^{1/2} + a_2 \theta + a_3 \theta^{3/2} + a_4 \theta^2 + \cdots \ .
\end{align}
With this ansatz, $\mu_{RN}$, $q_{RN}$, and $m_{RN}$ can be parametrized as
\begin{align}
\mu_{RN} &= \frac{2}{\sqrt{3} \kappa} + \frac{5 a_1^2}{3 \sqrt{3} \kappa} \theta + \frac{10 a_1 a_2}{3 \sqrt{3} \kappa} \theta^{3/2} + \frac{5 (-5 a_1^4 + 12 a_2^2 + 24 a_1 a_3)}{36 \sqrt{3} \kappa} \theta^2 \cr
& \qquad +  \frac{5 (-5 a_1^3 a_2 + 6 a_2 a_3 + 6 a_1 a_4)}{9 \sqrt{3} \kappa} \theta^{5/2} + \cdots \ , \cr
q_{RN} &= \frac{2 a_1^3}{\sqrt{3} \kappa} \theta^{3/2} + \frac{2 \sqrt{3} a_1^2 a_2}{\kappa} \theta^2 + \frac{5 a_1^5 + 18 a_1 a_2^2 + 18 a_1^2 a_3}{3 \sqrt{3} \kappa} \theta^{5/2} \cr
& \qquad + \frac{25 a_1^4 a_2 + 6 a_2^3 + 36 a_1 a_2 a_3 + 18 a_1^2 a_4}{3 \sqrt{3} \kappa} \theta^3 + \cdots \ , \cr
m_{RN} &= 2 a_1^3 \theta^{3/2} + 6 a_1^2 a_2 \theta^2 + \left( \frac{8 a_1^5}{3} + 6 a_1 a_2^2 + 6 a_1^2 a_3 \right) \theta^{5/2} \cr
& \qquad + \left( \frac{40 a_1^4 a_2}{3} + 2 a_2^3 + 12 a_1 a_2 a_3 + 6 a_1^4 a_4 \right) \theta^3 + \cdots \ .
\end{align}
For the balancing of the chemical potential, the ansatz for the charge of the soliton is chosen as
\begin{align}
q_{sol} = b_1 \theta + b_2 \theta^{3/2} + b_3 \theta^2 + b_4 \theta^{5/2} + \cdots \ .
\end{align}
The chemical potential and mass of the soliton are
\begin{align}
\mu_{sol} &= \frac{2}{\sqrt{3} \kappa} + \left( - \frac{1}{\sqrt{3}\kappa} + \frac{5 \times 2^{2-2\Delta} b_1 \, \Gamma(\Delta) \, \Gamma(2 \Delta - 5/2)}{3 \, \Gamma(\Delta+1/2) \, \Gamma(\Delta-3/2)^2} \right) \theta \cr
& \qquad + \frac{5 \times 2^{2-2\Delta} b_2 \, \Gamma(\Delta) \, \Gamma(2 \Delta - 5/2)}{3 \, \Gamma(\Delta+1/2) \, \Gamma(\Delta-3/2)^2} \theta^{3/2} \cr
& \qquad + \left( \frac{\sqrt{3}}{4 \kappa} +
 \frac{2^{2-2\Delta} (5 b_3 + 2 b_1 (2 \Delta - 5)) \Gamma(\Delta) \, \Gamma(2 \Delta - 5/2)}{3 \, \Gamma(\Delta+1/2) \, \Gamma(\Delta-3/2)^2} \right) \theta^2 + \cdots \ , \cr
m_{sol} &= \sqrt{3} \kappa b_1 \theta + \sqrt{3} \kappa b_2 \theta^{3/2} + \left( \frac{\sqrt{3}}{2} \kappa (2 b_3 - b_1) + \frac{5 b_1^2 \kappa^2 \, \Gamma(\Delta) \, \Gamma(2 \Delta - 5/2)}{2^{2 \Delta} \, \Gamma(\Delta+1/2) \, \Gamma(\Delta-3/2)^2} \right) \theta^2 \cr
& \qquad + \frac{1}{2} \left( \sqrt{3} \kappa (2 b_4 - b_2) +  \frac{5 b_1 b_2 \kappa^2 \, \Gamma(\Delta) \, \Gamma(2 \Delta - 5/2)}{2^{2\Delta-2} \, \Gamma(\Delta+1/2) \, \Gamma(\Delta-3/2)^2} \right) \theta^{5/2} + \cdots \ ,
\end{align}
where $b_1>0$. The charge of the hairy black hole is
\begin{align}
q = q_{RN} + q_{sol} = b_1 \theta + \left( \frac{2 a_1^3}{\sqrt{3} \kappa} + b_2 \right) \theta^{3/2} + \cdots \ .
\label{eq_qmix_d5_e>ec}
\end{align}
Solving $\mu_{RN} = \mu_{sol}$ and $q \sim \theta$, we obtain $a_{1,2,3,4}$ and $b_{1,2,3,4}$. The mass $m_{hairy} = m_{RN} + m_{sol}$ is then expressed as the following function of $q$, in $q>q_c$,
\begin{align}
m_{hairy} = m_{sol} - \frac{3^{1/4} \, \Gamma(\Delta)^{5/2} \, \Gamma(2\Delta-5/2)^{5/2}}{2^{5 \Delta - 6} \, \Gamma(\Delta-3/2)^5 \, \Gamma(\Delta+1/2)^{5/2}} \kappa^{5/2} (q-q_c)^{5/2} \ ,
\end{align}
where
\begin{align}
q_c &= \frac{2^{2 \Delta-2} \sqrt{3} \, \Gamma(\Delta-3/2)^2 \, \Gamma(\Delta+1/2)}{5 \kappa \, \Gamma(\Delta) \, \Gamma(2\Delta-5/2)} \theta \ , \cr
m_{sol} &= \sqrt{3} \kappa q + \left(\frac{5 \, \Gamma(\Delta) \, \Gamma(2\Delta-5/2)}{2^{2 \Delta} \, \Gamma(\Delta-3/2)^2 \, \Gamma(\Delta+1/2)} \kappa^2 q^2 - \frac{\sqrt{3}}{2} \kappa q \theta \right) \ .
\end{align}
In $q<q_c$, no real solution satisfying $\mu_{RN} = \mu_{sol}$ was obtained.


\begin{thebibliography}{99} 

\bibitem{Landau:1980mil}
L.~D.~Landau and E.~M.~Lifshitz,
Butterworth-Heinemann, 1980,
ISBN 978-0-7506-3372-7





\bibitem{Arkani-Hamed:2006emk}
N.~Arkani-Hamed, L.~Motl, A.~Nicolis and C.~Vafa,
JHEP \textbf{06}, 060 (2007)
doi:10.1088/1126-6708/2007/06/060
[arXiv:hep-th/0601001 [hep-th]].

\bibitem{Harlow:2022ich}
D.~Harlow, B.~Heidenreich, M.~Reece and T.~Rudelius,
Rev. Mod. Phys. \textbf{95}, no.3, 3 (2023)
doi:10.1103/RevModPhys.95.035003
[arXiv:2201.08380 [hep-th]].


\bibitem{Aharony:2021mpc}
O.~Aharony and E.~Palti,
Phys. Rev. D \textbf{104}, no.12, 126005 (2021)
doi:10.1103/PhysRevD.104.126005
[arXiv:2108.04594 [hep-th]].


\bibitem{Antipin:2021rsh}
O.~Antipin, J.~Bersini, F.~Sannino, Z.~W.~Wang and C.~Zhang,
JHEP \textbf{12}, 204 (2021)
doi:10.1007/JHEP12(2021)204
[arXiv:2109.04946 [hep-th]].


\bibitem{Moser:2021bes}
R.~Moser, D.~Orlando and S.~Reffert,
JHEP \textbf{02}, 152 (2022)
doi:10.1007/JHEP02(2022)152
[arXiv:2110.07617 [hep-th]].


\bibitem{Palti:2022unw}
E.~Palti and A.~Sharon,
JHEP \textbf{09}, 078 (2022)
doi:10.1007/JHEP09(2022)078
[arXiv:2206.06703 [hep-th]].


\bibitem{Orlando:2023ljh}
D.~Orlando and E.~Palti,
Phys. Rev. D \textbf{108}, no.8, 085002 (2023)
doi:10.1103/PhysRevD.108.085002
[arXiv:2303.02178 [hep-th]].



\bibitem{Aharony:2023ike}
O.~Aharony and Y.~N.~Breitstein,
JHEP \textbf{08}, 044 (2023)
doi:10.1007/JHEP08(2023)044
[arXiv:2305.08947 [hep-th]].



\bibitem{Sharon:2023drx}
A.~Sharon and M.~Watanabe,
JHEP \textbf{05}, 202 (2023)
doi:10.1007/JHEP05(2023)202
[arXiv:2301.08262 [hep-th]].




\bibitem{Nakayama:2015hga}
Y.~Nakayama and Y.~Nomura,
Phys. Rev. D \textbf{92}, no.12, 126006 (2015)
doi:10.1103/PhysRevD.92.126006
[arXiv:1509.01647 [hep-th]].


\bibitem{Basu:2010uz}
P.~Basu, J.~Bhattacharya, S.~Bhattacharyya, R.~Loganayagam, S.~Minwalla and V.~Umesh,
JHEP \textbf{10}, 045 (2010)
doi:10.1007/JHEP10(2010)045
[arXiv:1003.3232 [hep-th]].


\bibitem{Bhattacharyya:2010yg}
S.~Bhattacharyya, S.~Minwalla and K.~Papadodimas,
JHEP \textbf{11}, 035 (2011)
doi:10.1007/JHEP11(2011)035
[arXiv:1005.1287 [hep-th]].


\bibitem{Dias:2011tj}
O.~J.~C.~Dias, P.~Figueras, S.~Minwalla, P.~Mitra, R.~Monteiro and J.~E.~Santos,
JHEP \textbf{08}, 117 (2012)
doi:10.1007/JHEP08(2012)117
[arXiv:1112.4447 [hep-th]].


\bibitem{Maeda:2010hf}
K.~Maeda, S.~Fujii and J.~i.~Koga,
Phys. Rev. D \textbf{81}, 124020 (2010)
doi:10.1103/PhysRevD.81.124020
[arXiv:1003.2689 [gr-qc]].


\bibitem{Gentle:2011kv}
S.~A.~Gentle, M.~Rangamani and B.~Withers,
JHEP \textbf{05}, 106 (2012)
doi:10.1007/JHEP05(2012)106
[arXiv:1112.3979 [hep-th]].


\bibitem{Dias:2016pma}
\'O.~J.~C.~Dias and R.~Masachs,
JHEP \textbf{02}, 128 (2017)
doi:10.1007/JHEP02(2017)128
[arXiv:1610.03496 [hep-th]].


\bibitem{Basu:2016mol}
P.~Basu, C.~Krishnan and P.~N.~Bala Subramanian,
JHEP \textbf{06}, 139 (2016)
doi:10.1007/JHEP06(2016)139
[arXiv:1602.07211 [hep-th]].




\bibitem{Loukas:2018zjh}
O.~Loukas, D.~Orlando, S.~Reffert and D.~Sarkar,
Nucl. Phys. B \textbf{934}, 437-458 (2018)
doi:10.1016/j.nuclphysb.2018.07.020
[arXiv:1804.04151 [hep-th]].


\bibitem{Nakayama:2020dle}
Y.~Nakayama,
Phys. Lett. B \textbf{808}, 135677 (2020)
doi:10.1016/j.physletb.2020.135677
[arXiv:2004.08069 [hep-th]].




\bibitem{Giombi:2017mxl}
S.~Giombi and E.~Perlmutter,
JHEP \textbf{03}, 026 (2018)
doi:10.1007/JHEP03(2018)026
[arXiv:1709.09159 [hep-th]].


\bibitem{Urbano:2018kax}
A.~Urbano,
[arXiv:1810.05621 [hep-th]].



\bibitem{Montero:2018fns}
M.~Montero,
JHEP \textbf{03}, 157 (2019)
doi:10.1007/JHEP03(2019)157
[arXiv:1812.03978 [hep-th]].



\bibitem{Alday:2019qrf}
L.~F.~Alday and E.~Perlmutter,
JHEP \textbf{08}, 084 (2019)
doi:10.1007/JHEP08(2019)084
[arXiv:1906.01477 [hep-th]].



\bibitem{Cho:2023koe}
M.~Cho, S.~Choi, K.~H.~Lee and J.~Song,
JHEP \textbf{11}, 118 (2023)
doi:10.1007/JHEP11(2023)118
[arXiv:2308.01717 [hep-th]].


\bibitem{Crisford:2017gsb}
T.~Crisford, G.~T.~Horowitz and J.~E.~Santos,
Phys. Rev. D \textbf{97}, no.6, 066005 (2018)
doi:10.1103/PhysRevD.97.066005
[arXiv:1709.07880 [hep-th]].


\bibitem{Fitzpatrick:2011hh}
A.~L.~Fitzpatrick and D.~Shih,
JHEP \textbf{10}, 113 (2011)
doi:10.1007/JHEP10(2011)113
[arXiv:1104.5013 [hep-th]].


\bibitem{Andriolo:2022hax}
S.~Andriolo, M.~Michel and E.~Palti,
JHEP \textbf{02}, 078 (2023)
doi:10.1007/JHEP02(2023)078
[arXiv:2211.04477 [hep-th]].


\bibitem{Hawking:1999dp}
S.~W.~Hawking and H.~S.~Reall,
Phys. Rev. D \textbf{61}, 024014 (2000)
doi:10.1103/PhysRevD.61.024014
[arXiv:hep-th/9908109 [hep-th]].


\bibitem{Uchikata:2011zz}
N.~Uchikata and S.~Yoshida,
Phys. Rev. D \textbf{83}, 064020 (2011)
doi:10.1103/PhysRevD.83.064020
[arXiv:1109.6737 [gr-qc]].


\bibitem{Green:2015kur}
S.~R.~Green, S.~Hollands, A.~Ishibashi and R.~M.~Wald,
Class. Quant. Grav. \textbf{33}, no.12, 125022 (2016)
doi:10.1088/0264-9381/33/12/125022
[arXiv:1512.02644 [gr-qc]].


\bibitem{Hartnoll:2008vx}
S.~A.~Hartnoll, C.~P.~Herzog and G.~T.~Horowitz,
Phys. Rev. Lett. \textbf{101}, 031601 (2008)
doi:10.1103/PhysRevLett.101.031601
[arXiv:0803.3295 [hep-th]].


\bibitem{Hartnoll:2008kx}
S.~A.~Hartnoll, C.~P.~Herzog and G.~T.~Horowitz,
JHEP \textbf{12}, 015 (2008)
doi:10.1088/1126-6708/2008/12/015
[arXiv:0810.1563 [hep-th]].


\bibitem{Agarwal:2020zwm}
P.~Agarwal, S.~Choi, J.~Kim, S.~Kim and J.~Nahmgoong,
Phys. Rev. D \textbf{103}, no.12, 126006 (2021)
doi:10.1103/PhysRevD.103.126006
[arXiv:2005.11240 [hep-th]].


\bibitem{Choi:2021rxi}
S.~Choi, S.~Jeong, S.~Kim and E.~Lee,
JHEP \textbf{09}, 138 (2023)
doi:10.1007/JHEP09(2023)138
[arXiv:2111.10720 [hep-th]].


\bibitem{Hellerman:2015nra}
S.~Hellerman, D.~Orlando, S.~Reffert and M.~Watanabe,
JHEP \textbf{12}, 071 (2015)
doi:10.1007/JHEP12(2015)071
[arXiv:1505.01537 [hep-th]].



\bibitem{Jafferis:2017zna}
D.~Jafferis, B.~Mukhametzhanov and A.~Zhiboedov,
JHEP \textbf{05}, 043 (2018)
doi:10.1007/JHEP05(2018)043
[arXiv:1710.11161 [hep-th]].


\bibitem{delaFuente:2020yua}
A.~de la Fuente and J.~Zosso,
JHEP \textbf{06}, 178 (2020)
doi:10.1007/JHEP06(2020)178
[arXiv:2005.06169 [hep-th]].


\bibitem{Harada:2023cfl}
T.~Harada, T.~Ishii, T.~Katagiri and N.~Tanahashi,
JHEP \textbf{06}, 106 (2023)
doi:10.1007/JHEP06(2023)106
[arXiv:2304.02267 [hep-th]].


\bibitem{Witten:2001ua}
E.~Witten,
[arXiv:hep-th/0112258 [hep-th]].


\bibitem{Berkooz:2002ug}
M.~Berkooz, A.~Sever and A.~Shomer,
JHEP \textbf{05}, 034 (2002)
doi:10.1088/1126-6708/2002/05/034
[arXiv:hep-th/0112264 [hep-th]].




\bibitem{Cheung:2018cwt}
C.~Cheung, J.~Liu and G.~N.~Remmen,
JHEP \textbf{10}, 004 (2018)
doi:10.1007/JHEP10(2018)004
[arXiv:1801.08546 [hep-th]].



\bibitem{Hamada:2018dde}
Y.~Hamada, T.~Noumi and G.~Shiu,
Phys. Rev. Lett. \textbf{123}, no.5, 051601 (2019)
doi:10.1103/PhysRevLett.123.051601
[arXiv:1810.03637 [hep-th]].



\bibitem{Balasubramanian:1998sn}
V.~Balasubramanian, P.~Kraus and A.~E.~Lawrence,
Phys. Rev. D \textbf{59}, 046003 (1999)
doi:10.1103/PhysRevD.59.046003
[arXiv:hep-th/9805171 [hep-th]].


\bibitem{Breitenlohner:1982bm}
P.~Breitenlohner and D.~Z.~Freedman,
Phys. Lett. B \textbf{115}, 197-201 (1982)
doi:10.1016/0370-2693(82)90643-8


\bibitem{Breitenlohner:1982jf}
P.~Breitenlohner and D.~Z.~Freedman,
Annals Phys. \textbf{144}, 249 (1982)
doi:10.1016/0003-4916(82)90116-6

\end{thebibliography}
\end{document}